\documentclass{article}

\usepackage[numbers]{natbib}
\PassOptionsToPackage{numbers, compress}{natbib}
\usepackage[main, final]{neurips_2026}

\usepackage[utf8]{inputenc} 
\usepackage[T1]{fontenc}    
\usepackage{hyperref}       
\usepackage{url}            
\usepackage{booktabs}       
\usepackage{amsfonts}       
\usepackage{nicefrac}       
\usepackage{microtype}      
\usepackage{xcolor}         
\usepackage{multirow}
\usepackage{makecell}
\usepackage{booktabs}
\usepackage{graphicx}
\usepackage{colortbl}
\usepackage{amsmath}
\usepackage{float}
\usepackage{wrapfig}
\usepackage[ruled,vlined,linesnumbered]{algorithm2e}
\definecolor{bettercolor}{RGB}{200,240,200}
\definecolor{worsecolor}{RGB}{255,220,220}
\definecolor{neutralcolor}{RGB}{240,240,240}

\definecolor{bestcell}{RGB}{198,239,206}
\definecolor{warnCell}{RGB}{255,220,180}
\definecolor{badCell}{RGB}{255,199,206}
\definecolor{hdrBlue}{RGB}{221,235,247}

\newcommand{\bad}[1]{\cellcolor{badCell}{#1}}

\definecolor{goodcell}{RGB}{198,239,206}
\definecolor{medcell}{RGB}{255,235,156}
\definecolor{badcell}{RGB}{255,199,206}
\definecolor{deadcell}{RGB}{210,210,210}
\definecolor{sectiongray}{RGB}{242,242,242}
 
\newcommand{\good}[1]{\cellcolor{goodcell}{#1}}

\newcommand{\dead}[1]{\cellcolor{deadcell}{#1}}

\usepackage{array}
\title{Progressive Layer-wise Supervision:\\ Deep-to-Shallow Supervision Annealing for Efficient and Robust Speech Deepfake Detection}

\author{%
  Hoan My Tran \\
  Université de Rennes, CNRS, IRISA\\
  Lannion, France\\
  \texttt{hoan.tran@irisa.fr} \\
  \And
  Xin Wang \\
  National Institute of Informatics \\
  Tokyo, Japan \\
  \texttt{wangxin@nii.ac.jp} \\
  \AND
  Xuechen Liu \\
  Xi’an Jiaotong-Liverpool University \\
  Taicang, China \\
  \texttt{xuechen.liu@xjtlu.edu.cn} \\
}

\begin{document}

\maketitle
\vspace{-2em}
\begin{center}
    \href{https://github.com/hoanmyTran/pls-deepfake-detection}{\includegraphics[height=0.8em]{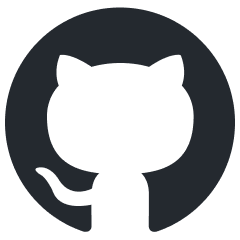} \textcolor{blue}{\texttt{GitHub}}} \quad
\end{center}

\begin{abstract}

Speech foundation models pretrained via self-supervised learning provide powerful representations for deepfake detection, yet we find that standard fine-tuning exploits only the deepest transformer layers — leaving the majority of the network's capacity discriminatively inert. Layer-wise probing of \textit{XLS-R-128} reveals that the top two-thirds of the transformer stack contribute negligibly to anti-spoofing performance after standard fine-tuning, a phenomenon we term shallow-layer dormancy. We connect this observation to the representational redundancy of pretrained transformers: adjacent layers encode highly similar information, and fine-tuning without explicit intermediate supervision reinforces rather than breaks this redundancy. To address this, we propose Progressive Layer-wise Supervision (PLS), a training framework that attaches a shared linear classifier to every transformer layer and schedules layer-wise loss contributions via an exponentially annealed power weighting. Early training concentrates supervision on deep layers to preserve the pretrained representational hierarchy; later training progressively redistributes supervision to shallower layers, converting dormant earlier representations into discriminative ones. Competitive detection ($\leq9.25\%$ average EER) is achievable on \emph{XLS-R-128} at layer 7, which reduces the parameter count from 315M to 101M and exemplifies an early exit mechanism. Experiments across twelve benchmarks spanning in-domain, and out-of-domain deepfake scenarios show PLS achieves a new state-of-the-art with 239M parameters, outperforming prior methods with larger and more complex architectures.
\end{abstract}

\section{Introduction}
\label{sec:introduction}
The rapid advancement of neural speech synthesis has made AI-generated voices increasingly indistinguishable from genuine recordings~\cite{tan2021survey}. Text-to-speech (TTS) and voice conversion (VC) technologies, while enabling beneficial applications~\cite{medeiros2015intel}, are increasingly misused for impersonation, fraudulent robocalls, and identity theft~\cite{folorunsho2025deepfake}. Detecting synthetic speech has become a pressing forensic and societal challenge. State-of-the-art (SOTA) countermeasures leverage speech foundation models (SFMs) — large transformers pretrained on vast speech corpora via self-supervised learning (SSL) — as front-end feature extractors~\cite{tak22_odyssey, zhu2024slim, zhang2024audio, truong24b_interspeech, tran25b_interspeech, 11226534, 10.1145/3746027.3754568}. These models are subsequently fine-tuned for the binary \emph{bona fide} (real) / \emph{spoofed} (fake) classification task. Despite strong in-domain performance, they generalize poorly to out-of-domain (OOD) scenarios involving unseen synthesizers~\cite{muller22_interspeech, 10.1145/3543507.3583222, 10208955, 10832250, 10650962, WANG2026101825, 10839331, 10889450}, limiting their real-world utility.

\paragraph{The shallow-layer dormancy problem.} We identify a specific and underexplored failure mode of standard SFM fine-tuning. Layer-wise probing of \emph{XLS-R-128} 300M (Figure~\ref{fig:proposed_baseline_equal}, Table~\ref{tab:xlsr300m_standard_aligned}) reveals that layers 0–15 of 24 consistently yield average equal error rate (EER) exceeding 20\% across all twelve evaluation benchmarks after standard fine-tuning — competitive performance (EER $< 12\%$) emerges only from layer 16 onward. Two-thirds of the network contributes nothing to deepfake detection, despite containing over 200M parameters. This observation is not merely about efficiency. We hypothesize that these dormant layers represent untapped generalization capacity: if earlier representations could be made discriminative, they would provide the model with multiple independent detection pathways, reducing its brittle dependence on the final-layer representation and improving robustness to distribution shift.

\paragraph{Connection to representational redundancy.} Prior work has extensively documented that pretrained SFMs exhibit high representational similarity across adjacent layers~\cite{9688093, 10096149, 10887866}. This redundancy arises from SSL pretraining objectives that optimize for smooth, hierarchical speech representations — not for task-specific discrimination. Standard fine-tuning, which applies supervision only at the final layer, fails to disrupt this redundancy in earlier layers: they remain anchored to the pretrained representational geometry~\cite{papyan2020prevalence, gromov2025the, jiang2025tracing, sukenik2025neural}, which was not designed for anti-spoofing. Explicit supervision is necessary to realign these layers toward the detection task.

\paragraph{The ordering problem.} A response is to supervise all layers simultaneously with uniform weights. We show empirically, however, this is counterproductive: applying strong supervision to shallow layers before deep layers have adapted to the anti-spoofing task degrades final-layer performance by 1.2–3.5\% EER. This is because the pretrained representational hierarchy is valuable — high-level semantic and acoustic features are organized progressively across depth~\cite{9688093, 10096149, 10887866}, and disrupting shallow layers before deep layers have stabilized collapses this structure prematurely~\cite{jiang2025tracing, sukenik2025neural}.

\paragraph{Our approach.} We propose Progressive Layer-wise Supervision (PLS), which supervises all layers through a shared linear classifier while scheduling layer-wise loss contributions via an exponentially annealed power weighting. The schedule concentrates supervision on deep layers early in training, allowing high-level representations to stabilize first, then progressively redistributes supervision toward shallower layers, awakening dormant earlier representations in a controlled, bottom-up order. An even/odd epoch alternation further anchors the primary classification objective, preventing drift from accumulated intermediate supervision. We make the following specific contributions:

\begin{itemize}
    \item \textbf{Progressive Layer-wise Supervision (PLS).} We introduce an exponentially annealed power weighting for layer-wise losses that activates the full transformer stack in a deep-to-shallow order, converting redundant earlier representations into discriminative ones.
    \item \textbf{Efficient early-exit inference.} The shared classifier enables inference to exit at layer 7 of 24, reducing parameters by 70\% with competitive performance under 10\% of average EER.
    \item \textbf{State-of-the-art OOD generalization.} PLS achieves 6.03\% average EER across eight datasets, outperforming all prior methods with larger and more complex backends.
\end{itemize}

\section{Related work}
\label{sec:related_work}

\paragraph{Deepfake generation and anti-spoofing countermeasures.}
 
Recent advances in TTS and VC have enabled the synthesis of speech that is nearly indistinguishable from genuine human recordings~\cite{tan2021survey}. Modern TTS systems typically follow a multi-stage pipeline: textual input is converted into linguistic representations, which are then mapped to acoustic features and synthesized into waveforms using neural vocoders. VC systems preserve the linguistic content of a source utterance while transforming its speaker characteristics to match a target speaker, often from reference speech. Together, these technologies raise both opportunities and security concerns~\cite{10832320, ju2024naturalspeech, chen-etal-2025-f5}.

Early countermeasures relied on handcrafted features such as CQCCs~\cite{TODISCO2017516} and MFCCs~\cite{lei09c_interspeech} with classical classifiers. The field shifted to end-to-end neural architectures~\cite{tak21_asvspoof, 9747766, chen24k_interspeech}, and more recently, to SFM-based pipelines using \emph{Wav2Vec2}~\cite{baevski2020wav2vec} or \emph{WavLM}~\cite{chen2022wavlm} as front-end extractors followed by increasingly complex backends — graph attention networks~\cite{9747766}, Transformer~\cite{truong24b_interspeech}, Mamba~\cite{tran25b_interspeech}, Kolmogorov–Arnold networks~\cite{11226534, 11460320}, and feature gating~\cite{10.1145/3746027.3754568}. We improve how the SFM encoder itself is trained, and achieve competitive results with a simple linear backend.

\paragraph{Representational redundancy in pretrained speech models.}
Pretrained transformer encoders exhibit high representational similarity between adjacent layers~\cite{9688093, 10096149, 10887866}. \citet{9688093, 10096149} show via probing that SFMs encode phonetic and semantic information in a graduated hierarchy, with considerable redundancy particularly in middle layers. \citet{10887866} show that substantial portions of the transformer stack in SFMs can be pruned with minimal performance loss, confirming functional redundancy. This redundancy is connected theoretically to layer-wise saturation in deep networks~\cite{sukenik2025neural, jiang2025tracing, gromov2025the} and to the neural collapse phenomenon~\cite{papyan2020prevalence}, where class-conditional representations converge toward an equiangular tight frame geometry at the output layer. Our work leverages this redundancy constructively: rather than pruning redundant layers, we activate them for the downstream detection task that they can be pruned after training.

\paragraph{Multi-exit and layer-wise training.}
Multi-exit inference has been explored in NLP~\cite{xin-etal-2020-deebert, zhou2020bert}, where auxiliary classifiers at intermediate layers enable adaptive computation. ~\citet{jiang2025tracing} show in vision transformers that shared classifiers trained across all layers with aligned objectives can reduce representational redundancy and improve multi-exit performance. For audio deepfake detection, \citet{10810699} attach classifier at fixed intermediate layers of \emph{WavLM} for early exit. Our approach uses a shared classifier across all layers, which is both parameter-efficient and encourages representational alignment across depth while introducing the annealing schedule that we show is essential for preserving pretrained representational structure while activating intermediate layers.

\section{Method}
\label{sec:method}

\begin{wrapfigure}{r}{0.51\textwidth}
    \centering
    \vspace{-1.5em}
    \includegraphics[width=0.46\textwidth]{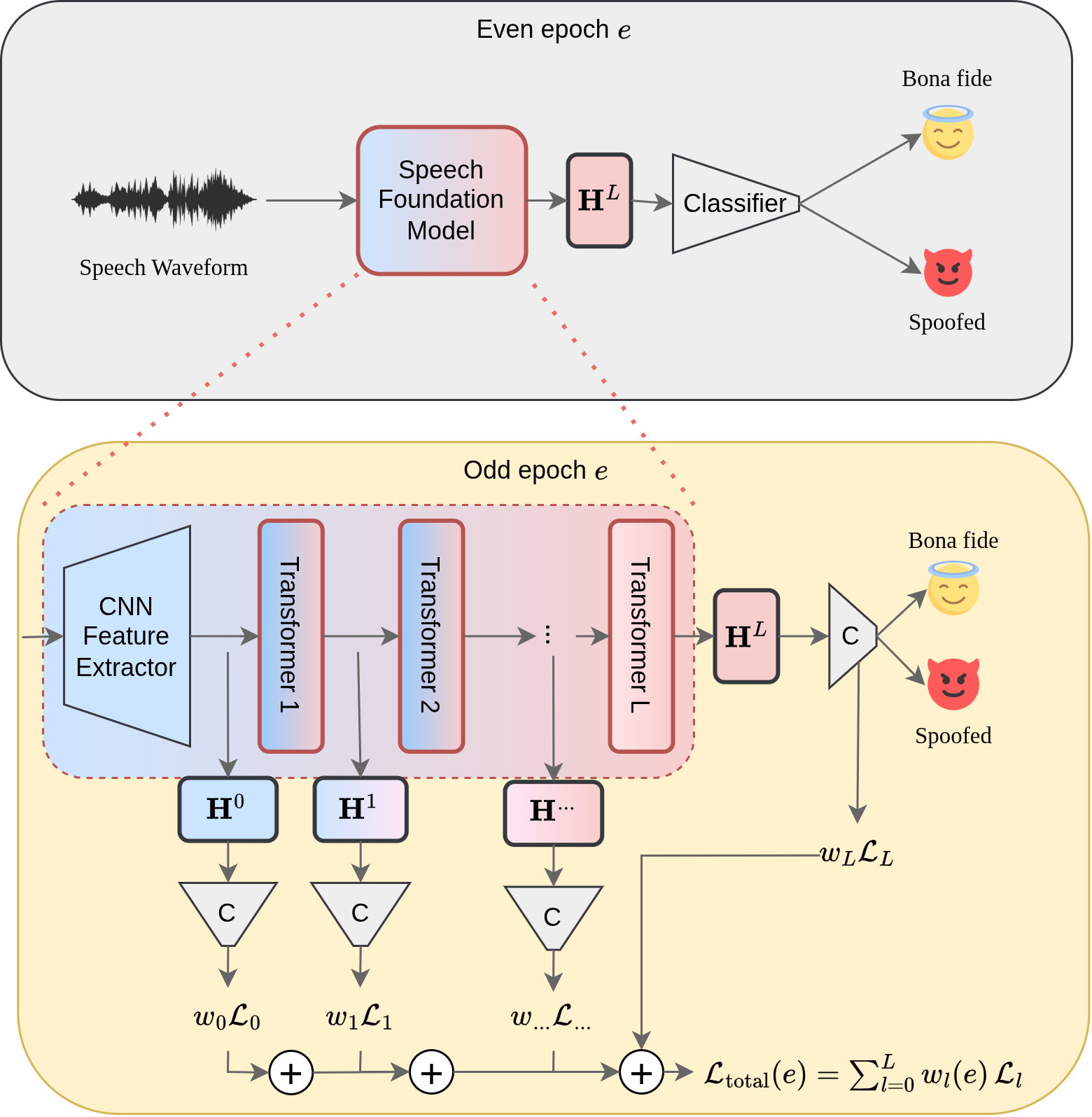}
    \caption{PLS training objective. During even epochs, the model is anchored by standard supervision on the final transformer layer $\mathbf{H}^{(L)}$. In odd epochs, a deep-to-shallow supervision is applied, where the total loss $\mathcal{L}_{\text{total}}$ is a weighted sum of all layer-wise losses $\mathcal{L}_l$. The layer weights $w_l$ evolve via a decaying power parameter $\rho(e)$, transitioning from high-level to uniform representation.}
    \label{fig:architecture}
\end{wrapfigure}

\paragraph{Preliminaries.}

SFMs consist of a convolutional feature extractor $f: \mathcal{X} \rightarrow \mathcal{Z}$ mapping raw waveform to latent frames $\mathcal{Z} = \{\mathbf{z}_1, \dots, \mathbf{z}_T\}$, followed by $L$ transformer layers. 
Let $\mathbf{H}^{(l)} = \{\mathbf{h}_1^{(l)}, \dots, \mathbf{h}_T^{(l)}\}$ denote hidden states at layer $l \in \{0, \dots, L\}$, where $l=0$ corresponds to the output of the convolutional feature extractor and $l \in \{1, \dots, L\}$ corresponds to the outputs of the subsequent transformer layers. Global average pooling over $\mathbf{H}^{(L)}$ yields a fixed-dimensional embedding; a shared linear classifier $c$ produces logits $\hat{\mathbf{y}}^{(l)}$ with binary cross-entropy loss $\mathcal{L}_L$.

\paragraph{Characterizing shallow-layer dormancy.}
We begin with an empirical characterization of the problem. After standard fine-tuning of \emph{XLS-R-128} 300M, we attach the trained linear classifier to each layer separately and evaluate on all twelve benchmarks (Table~\ref{tab:xlsr300m_standard_aligned}). When classifier head is attached to layers 0–7, it yields average EERs above 40\% on every benchmark, including in-domain sets. Competitive detection (EER < 12\%) emerges only from layer 16 onward, with the optimum at layer 21 (6.13\% Avg. EER). The shallow 15 of the 24 transformer layers, i.e., $l<16$, carry negligible discriminative information for anti-spoofing.

\paragraph{Why does standard fine-tuning fail to activate earlier layers?} 
While this is not surprising, we argue that the answer lies in representational redundancy across the layers and the way of fine-tuning.
SSL pretraining minimizes objectives that reward smooth, consistent representations across adjacent layers — producing representations highly similar to each other~\cite{9688093, 10096149}. 
In standard fine-tuning, gradients originate exclusively from the final classifier and decay through successive transformer blocks. Consequently, deeper layers receive stronger task-specific adaptation signals, while earlier layers remain close to their pretrained state. This produces a representational bottleneck in which discriminative information is concentrated near the output layers, leaving a substantial fraction of the network effectively underutilized for downstream deepfake detection task.

\begin{wrapfigure}{r}{0.6\textwidth} 
  \centering
  \includegraphics[width=0.6\textwidth]{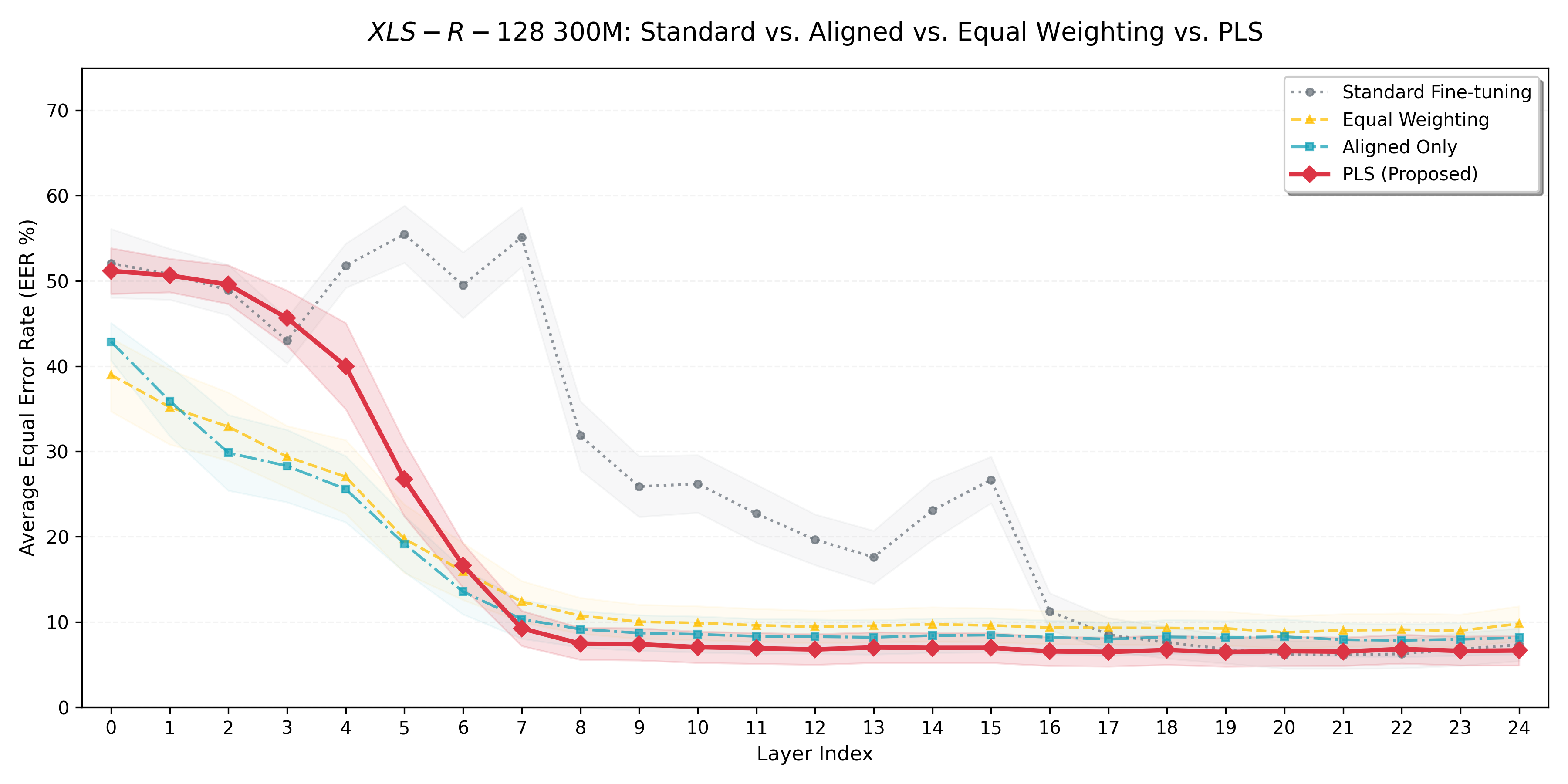}
  \caption{Proposed PLS approach comparing to standard, aligned, and equal weighting fine-tuning.}
  \label{fig:proposed_baseline_equal}
\end{wrapfigure}

\paragraph{Is uniform multi-layer supervision insufficient?}
The immediate remedy appears to be applying supervision to all layers simultaneously with uniform weights from the first epoch. We show empirically in Figure~\ref{fig:proposed_baseline_equal} that this is counterproductive: final-layer EER degrades by 2.2\% relative to standard fine-tuning, despite improving early-layer EER. 
We attribute this to pretraining structure disruption. The pretrained representational hierarchy — where progressively deeper layers encode progressively more abstract speech properties — is valuable for the downstream task~\cite{yang21c_interspeech, shi24g_interspeech}. When strong supervision is applied to shallow layers before the model has adapted its deeper representations to anti-spoofing, the resulting gradient updates propagate through the full stack and destabilize the deep-layer representations that are still in the process of being fine-tuned. Ablation study is found in Appendix~\ref{sec:supplementary_results} (Table~\ref{tab:xlsr_equal_weighting}).

\begin{figure}[htbp]
    \centering
    \includegraphics[width=\linewidth]{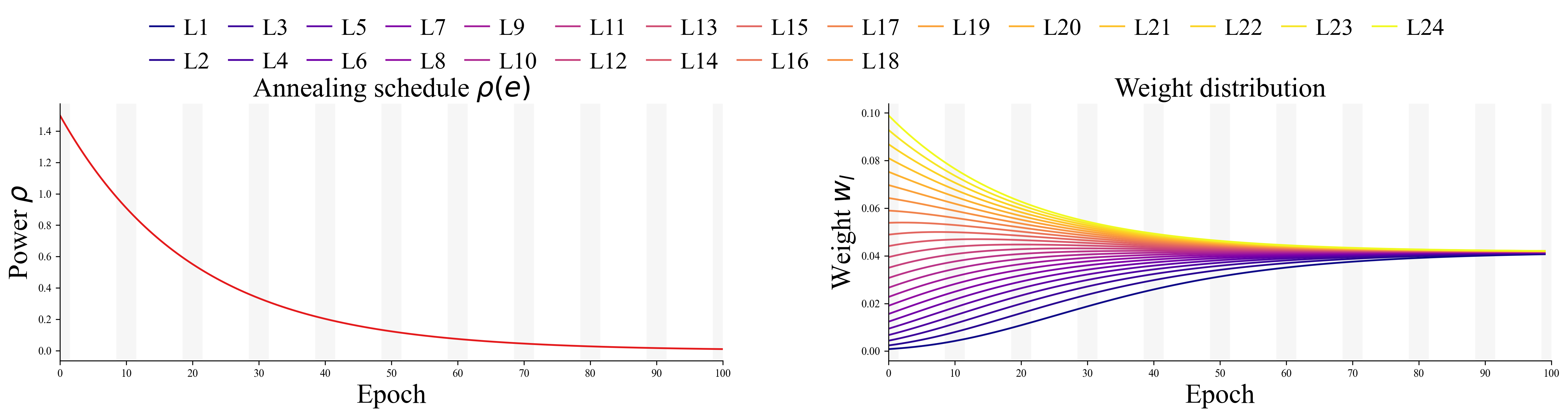}
    \caption{Illustration of annealing schedule, and weight distribution through epochs.}
    \label{fig:weights_gradients}
\end{figure}

\paragraph{Progressive Layer-wise Supervision.}
We define a scheduled multi-layer training objective. 
Let $\mathcal{L}_l$ denote the binary cross-entropy loss when the classifier head is attached to the $l$-th layer. At epoch $e$, the total training loss is:
\begin{equation}
    \label{eq:1}
    \mathcal{L}_{\text{total}}(e) = \sum_{l=0}^{L} w_l(e)\, \mathcal{L}_l.
\end{equation}

Layer weights are defined by an epoch-dependent power parameter $\rho(e)$:
\begin{equation}
    \label{eq:2}
    w_l(e) = \frac{(l+1)^{\rho(e)}}{\sum_{j=0}^{L} (j+1)^{\rho(e)}},
\end{equation}
where $\rho(e) = M \cdot \exp(-k e)$, with $M > 0$ controlling initial emphasis on deep layers and $k > 0$ controlling the annealing rate. PLS can be interpreted as encouraging a more gradual redistribution of task information across the transformer hierarchy, rather than concentrating discriminative capacity exclusively in the final layers.

\paragraph{Training dynamics.}
This schedule induces two natural phases (Figure~\ref{fig:weights_gradients}):
\begin{itemize}
    \item \textbf{Stabilization phase (large $\rho$, early epochs)}: Weights concentrate on $l \approx L$, The objective reduces approximately to standard final-layer fine-tuning, allowing deep representations to stabilize for the anti-spoofing task while shallow layers are only lightly updated.
    \item \textbf{Activation phase ($\rho \to 0$, late epochs)}: Weights converge toward $w_l \to \frac{1}{L+1}$, distributing supervision uniformly across all layers. Intermediate and shallow layers now receive direct supervision signals and progressively acquire discriminative structure.
\end{itemize}

\paragraph{Epoch alternation for primary objective anchoring.}
Even with annealing, accumulated multi-layer supervision over many odd-numbered epochs can drift the primary classification objective. To prevent this, we adopt an alternating strategy~\cite{jiang2025tracing}: for even-numbered epochs, we restrict supervision to the final layer only ($\mathcal{L}_{\text{total}}(e) = \mathcal{L}_L$), ensuring the primary classification objective remains well-anchored, as shown in Figure~\ref{fig:architecture}; for odd-numvered epochs, we apply the full scheduled objective $\mathcal{L}_{\text{total}}(e)$ from Equation~\ref{eq:1}.
The complete training algorithm is listed in Appendix~\ref{app:algorithm}.

\section{Experimental setup}
\label{sec:exp_setup}

All models are trained exclusively on the \texttt{ASVspoof 2019} logical access (\texttt{19LA}) training partition~\cite{wang2020asvspoof}, with model selection on the \texttt{19LA} development set. The \texttt{19LA} corpus contains bona fide speech from \texttt{VCTK}~\cite{yamagishi2019cstr} and spoofed samples generated by 17 TTS/VC systems — 6 seen during training, 13 reserved for evaluation. Weighted cross-entropy addresses class imbalance. 
RawBoost augmentation~\cite{tak2022rawboost} is applied. Audio inputs are dynamically padded. 

In Appendix~\ref{sec:impact_training_data}, we also trained the models on other corpora to evaluate the impact of the training dataset~\cite{WANG2026101825,10839331,huang-etal-2025-speechfake}. Detailed dataset descriptions and statistics are provided in Appendix~\ref{sec:appendix_dataset}.

For OOD evaluation, we use a diverse suite of datasets. \texttt{ASVspoof5} (\texttt{25DF})~\cite{WANG2026101825} contains utterances synthesized with advanced TTS/VC systems and novel adversarial attacks (ATs). \texttt{In-the-Wild} (\texttt{ITW})~\cite{muller22_interspeech} contains audio collected from social media platforms, \texttt{Fake-or-Real} (\texttt{FoR})~\cite{8906599} comprises samples from diverse open datasets and real-world recordings, and
\texttt{SpoofCeleb} (\texttt{SC})~\cite{10839331} includes TTS-generated speech with diverse speakers under noisy and reverberant conditions. Robustness to vocoder artefacts is assessed via \texttt{LibriSeVoc} (\texttt{LSV})~\cite{10208955}, which covers autoregressive, diffusion-based, and GAN-based vocoders. Diffusion and flow-matching scenarios are further evaluated on \texttt{DFADD}~\cite{10832250} and \texttt{DIFSSD}~\cite{10889450}. Finally, we also evaluated on \texttt{DECRO}~\cite{10.1145/3543507.3583222} and \texttt{MLAAD}~\cite{10650962}. All results are reported as EER (\%); the lower, the better. The code are available here\footnote{\href{https://github.com/TakHemlata/RawBoost-antispoofing}{RawBoost repository}}.
 
\begin{table}[htbp]
\centering
\caption{%
  \emph{XLS-R-128} 300M under \emph{standard} and \emph{aligned} fine-tuning: EER (\%, $\downarrow$) at each transformer; Avg. is the unweighted mean across 12 datasets.
}
\label{tab:xlsr300m_standard_aligned}
\resizebox{0.65\linewidth}{!}{%
\begin{tabular}{cccccccccccccccc}
\toprule
\multirow{2}{*}{\textbf{Layer}} &
\multirow{2}{*}{\textbf{Params.}} &
\multicolumn{3}{c}{\textbf{In-Domain}} &
\textbf{Voc.} &
\multicolumn{3}{c}{\textbf{In-the-Wild}} &
\multicolumn{2}{c}{\textbf{Diff/Flow}} &
\multicolumn{3}{c}{\textbf{Other OOD}} &
\multicolumn{2}{c}{\textbf{Summary}} \\
\cmidrule(lr){3-5}\cmidrule(lr){6-6}\cmidrule(lr){7-9}\cmidrule(lr){10-11}\cmidrule(lr){12-14}\cmidrule(lr){15-16}
& & \texttt{19LA} & \texttt{21LA} & \texttt{21DF} & \texttt{LSV}
& \texttt{ITW} & \texttt{FoR} & \texttt{SC}
& \texttt{DFADD} & \texttt{DIFSSD}
& \texttt{MLAAD-EN} & \texttt{DECRO-EN} & \texttt{25DF}
& \textbf{Avg.} & \textbf{Pool.} \\
\midrule
\multicolumn{16}{c}{\centering \textbf{Standard Fine-tuning}} \\
\midrule
\bad{0}  & \bad{13.13M} & \bad{69.59} & \bad{56.03} & \bad{58.48} & \bad{52.93} & \bad{15.34} & \bad{47.48} & \bad{52.91} & \bad{58.94} & \bad{43.23} & \bad{45.20} & \bad{73.13} & \bad{51.46} & \bad{52.06} & \bad{50.64} \\
\bad{1} & \bad{25.72M} & \bad{45.19} & \bad{38.82} & \bad{47.51} & \bad{52.59} & \bad{67.20} & \bad{44.00} & \bad{56.63} & \bad{45.96} & \bad{51.95} & \bad{31.70} & \bad{66.58} & \bad{61.33} & \bad{50.79} & \bad{45.00} \\
\bad{2} & \bad{38.32M} & \bad{47.90} & \bad{41.00} & \bad{46.62} & \bad{50.62} & \bad{64.22} & \bad{46.68} & \bad{50.64} & \bad{40.63} & \bad{44.08} & \bad{28.62} & \bad{68.07} & \bad{57.97} & \bad{48.92} & \bad{45.95} \\
\bad{3} & \bad{50.92M} & \bad{32.37} & \bad{36.16} & \bad{34.92} & \bad{45.02} & \bad{61.21} & \bad{53.09} & \bad{49.72} & \bad{38.80} & \bad{39.09} & \bad{32.28} & \bad{37.79} & \bad{55.44} & \bad{42.99} & \bad{43.92} \\
\bad{4} & \bad{63.51M} & \bad{46.00} & \bad{41.66} & \bad{42.01} & \bad{47.33} & \bad{72.86} & \bad{51.55} & \bad{54.17} & \bad{61.71} & \bad{56.28} & \bad{45.51} & \bad{43.50} & \bad{58.82} & \bad{51.78} & \bad{48.28} \\
\bad{5} & \bad{76.11M} & \bad{46.43} & \bad{42.26} & \bad{51.47} & \bad{51.87} & \bad{72.58} & \bad{74.38} & \bad{60.32} & \bad{64.27} & \bad{62.87} & \bad{37.66} & \bad{41.73} & \bad{59.83} & \bad{55.47} & \bad{53.47} \\
\bad{6} & \bad{88.71M} & \bad{38.95} & \bad{31.09} & \bad{41.09} & \bad{49.56} & \bad{75.34} & \bad{49.33} & \bad{64.25} & \bad{55.07} & \bad{60.39} & \bad{31.30} & \bad{37.64} & \bad{60.05} & \bad{49.51} & \bad{48.54} \\
\bad{7} & \bad{101.30M} & \bad{45.94} & \bad{44.69} & \bad{53.85} & \bad{51.65} & \bad{71.23} & \bad{72.44} & \bad{62.34} & \bad{49.14} & \bad{66.25} & \bad{38.95} & \bad{36.72} & \bad{68.30} & \bad{55.13} & \bad{58.95} \\
\midrule
\dead{8} & \dead{113.90M} & \dead{11.76} & \dead{13.13} & \dead{14.24} & \dead{23.25} & \dead{47.92} & \dead{44.78} & \dead{51.29} & \dead{34.17} & \dead{47.06} & \dead{41.39} & \dead{22.07} & \dead{30.96} & \dead{31.84} & \dead{29.13} \\
\dead{9} & \dead{126.49M} & \dead{15.64} & \dead{15.04} & \dead{ 9.13} & \dead{ 9.99} & \dead{33.59} & \dead{20.33} & \dead{37.26} & \dead{42.39} & \dead{42.07} & \dead{41.78} & \dead{25.71} & \dead{17.67} & \dead{25.88} & \dead{20.65} \\
\dead{10} & \dead{139.09M} & \dead{18.09} & \dead{17.45} & \dead{10.29} & \dead{11.70} & \dead{37.35} & \dead{16.62} & \dead{42.73} & \dead{36.69} & \dead{39.42} & \dead{40.15} & \dead{25.83} & \dead{18.01} & \dead{26.19} & \dead{21.63} \\
\dead{11} & \dead{151.69M} & \dead{13.50} & \dead{12.08} & \dead{ 6.93} & \dead{ 9.09} & \dead{28.22} & \dead{10.64} & \dead{34.98} & \dead{41.86} & \dead{31.45} & \dead{35.51} & \dead{31.38} & \dead{16.80} & \dead{22.70} & \dead{16.23} \\
\dead{12} & \dead{164.28M} & \dead{ 7.26} & \dead{ 9.53} & \dead{ 5.76} & \dead{ 8.67} & \dead{28.88} & \dead{10.55} & \dead{36.40} & \dead{28.72} & \dead{24.87} & \dead{31.19} & \dead{23.34} & \dead{20.57} & \dead{19.65} & \dead{17.49} \\
\dead{13} & \dead{176.88M} & \dead{ 3.70} & \dead{ 4.77} & \dead{ 4.77} & \dead{10.56} & \dead{29.25} & \dead{12.58} & \dead{34.57} & \dead{21.98} & \dead{31.98} & \dead{28.20} & \dead{11.84} & \dead{16.97} & \dead{17.60} & \dead{16.24} \\
\dead{14} & \dead{189.48M} & \dead{ 6.77} & \dead{ 9.05} & \dead{ 8.69} & \dead{16.96} & \dead{35.83} & \dead{17.67} & \dead{44.55} & \dead{28.47} & \dead{37.24} & \dead{34.19} & \dead{16.79} & \dead{20.82} & \dead{23.09} & \dead{20.11} \\
\dead{15} & \dead{202.07M} & \dead{18.83} & \dead{19.51} & \dead{11.40} & \dead{17.38} & \dead{31.95} & \dead{21.26} & \dead{33.95} & \dead{35.88} & \dead{37.53} & \dead{40.61} & \dead{34.00} & \dead{17.65} & \dead{26.66} & \dead{20.35} \\
\midrule
\good{16} & \good{214.67M} & \good{ 1.55} & \good{ 2.76} & \good{ 2.93} & \good{ 5.57} & \good{15.91} & \good{ 6.09} & \good{22.64} & \good{14.44} & \good{21.93} & \good{19.62} & \good{ 6.76} & \good{14.54} & \good{11.23} & \good{12.88} \\
\good{17} & \good{227.26M} &\good{ 1.17} & \good{ 2.96} & \good{ 1.88} & \good{ 2.30} & \good{ 8.29} & \good{ 2.44} & \good{18.73} & \good{12.43} & \good{16.73} & \good{18.74} & \good{ 3.30} & \good{13.35} & \good{8.53} & \good{12.38} \\
\good{18} & \good{239.86M} & \good{ 0.73} & \good{ 2.42} & \good{ 1.77} & \good{ 2.01} & \good{ 6.40} & \good{ 1.99} & \good{18.74} & \good{ 9.53} & \good{14.90} & \good{17.06} & \good{ 2.25} & \good{13.07} & \good{7.57} & \good{12.31} \\
\good{19} & \good{252.46M} & \good{ 0.34} & \good{ 2.49} & \good{ 1.53} & \good{ 2.28} & \good{ 5.69} & \good{ 2.35} & \good{17.97} & \good{ 8.62} & \good{12.34} & \good{14.56} & \good{ 1.16} & \good{12.53} & \good{6.82} & \good{12.20} \\
\good{20} & \good{265.05M} & \good{ 0.11} & \good{ 2.11} & \good{ 1.41} & \good{ 1.67} & \good{ 5.22} & \good{ 1.23} & \good{17.63} & \good{ 7.13} & \good{10.33} & \good{12.80} & \good{ 1.39} & \good{12.87} & \good{6.16} & \good{12.12} \\
\good{21} & \good{277.65M} & \good{ 0.07} & \good{ 2.07} & \good{ 1.42} & \good{ 1.67} & \good{ 5.09} & \good{ 1.01} & \good{17.59} & \good{ 6.89} & \good{10.63} & \good{11.57} & \good{ 2.32} & \good{13.18} & \good{5.80} & \good{11.99} \\
\good{22} & \good{290.25M} &\good{ 0.07} & \good{ 2.22} & \good{ 1.43} & \good{ 1.70} & \good{ 4.82} & \good{ 1.01} & \good{18.65} & \good{ 6.63} & \good{10.51} & \good{11.91} & \good{ 2.13} & \good{14.01} & \good{6.26} & \good{11.74} \\
\good{23} & \good{302.84M} & \good{ 0.10} & \good{ 2.31} & \good{ 1.49} & \good{ 1.67} & \good{ 4.59} & \good{ 0.88} & \good{20.62} & \good{ 7.81} & \good{10.24} & \good{12.31} & \good{ 1.86} & \good{17.81} & \good{6.81} & \good{11.99} \\
\good{24} & \good{315.44M} & \good{ 0.10} & \good{ 2.36} & \good{ 1.46} & \good{ 1.75} & \good{ 4.69} & \good{ 0.93} & \good{19.57} & \good{17.51} & \good{10.16} & \good{12.44} & \good{ 3.48} & \good{13.47} & \good{7.33} & \good{11.70} \\
\midrule
\multicolumn{16}{c}{\centering \textbf{Aligned Fine-tuning}} \\
\midrule
\bad{0} & \bad{13.13M}  & \bad{21.98} & \bad{22.08} & \bad{35.17} & \bad{53.23} & \bad{49.12} & \bad{82.24} & \bad{49.20} & \bad{23.57} & \bad{54.53} & \bad{34.74} & \bad{23.37} & \bad{50.51} & \bad{41.65} & \bad{43.52} \\
\bad{1} & \bad{25.72M} & \bad{31.84} & \bad{31.85} & \bad{34.56} & \bad{55.51} & \bad{45.51} & \bad{80.25} & \bad{44.75} & \bad{22.28} & \bad{47.42} & \bad{46.41} & \bad{27.38} & \bad{52.08} & \bad{43.32} & \bad{51.35} \\
\midrule
\dead{2} & \dead{38.32M} & \dead{24.16} & \dead{27.40} & \dead{28.89} & \dead{48.88} & \dead{34.11} & \dead{83.92} & \dead{42.20} & \dead{17.62} & \dead{39.30} & \dead{41.14} & \dead{20.06} & \dead{49.29} & \dead{38.08} & \dead{45.60} \\
\dead{3} & \dead{50.92M} &\dead{20.16} & \dead{24.84} & \dead{25.02} & \dead{45.77} & \dead{26.48} & \dead{73.20} & \dead{40.90} & \dead{20.00} & \dead{40.16} & \dead{40.68} & \dead{16.84} & \dead{42.10} & \dead{34.68} & \dead{41.65} \\
\dead{4} & \dead{63.51M} & \dead{19.17} & \dead{19.20} & \dead{19.92} & \dead{32.97} & \dead{24.41} & \dead{50.00} & \dead{43.29} & \dead{18.39} & \dead{35.50} & \dead{36.08} & \dead{16.79} & \dead{42.73} & \dead{29.87} & \dead{37.96} \\
\dead{5} & \dead{76.11M} & \dead{7.08}  & \dead{9.82}  & \dead{10.73} & \dead{11.66} & \dead{15.39} & \dead{22.23} & \dead{35.25} & \dead{13.37} & \dead{39.15} & \dead{24.54} & \dead{7.94}  & \dead{46.32} & \dead{20.29} & \dead{33.46} \\
\dead{6} & \dead{88.71M} & \dead{1.17}  & \dead{6.03}  & \dead{3.52}  & \dead{3.33}  & \dead{8.40}  & \dead{7.90}  & \dead{23.89} & \dead{11.76} & \dead{18.97} & \dead{20.42} & \dead{3.72}  & \dead{41.16} & \dead{12.52} & \dead{27.81} \\
\midrule
\good{7}  & \good{101.30M} & \good{0.29} & \good{3.96} & \good{1.94} & \good{1.82} & \good{4.87} & \good{3.22} & \good{19.95} & \good{7.94} & \good{17.96} & \good{14.71} & \good{2.53} & \good{35.58} & \good{9.56} & \good{23.80} \\
\good{8}  & \good{113.90M} & \good{0.23} & \good{3.91} & \good{1.86} & \good{1.14} & \good{4.12} & \good{2.52} & \good{19.33} & \good{6.38} & \good{14.07} & \good{16.10} & \good{1.33} & \good{21.63} & \good{7.72}  & \good{16.07} \\
\good{9} & \good{126.49M} & \good{0.11} & \good{3.62} & \good{1.76} & \good{0.99} & \good{4.07} & \good{2.33} & \good{18.69} & \good{6.91} & \good{13.80} & \good{15.28} & \good{1.35} & \good{22.24} & \good{7.60}  & \good{16.36} \\
\good{10} & \good{139.09M} & \good{0.10} & \good{3.38} & \good{1.67} & \good{1.02} & \good{4.00} & \good{2.07} & \good{18.58} & \good{7.03} & \good{13.89} & \good{15.00} & \good{1.55} & \good{21.23} & \good{7.46}  & \good{15.76} \\
\good{11} & \good{151.69M} & \good{0.07} & \good{3.02} & \good{1.60} & \good{0.95} & \good{3.97} & \good{2.11} & \good{18.57} & \good{7.13} & \good{13.51} & \good{14.09} & \good{1.51} & \good{24.56} & \good{7.59}  & \good{18.23} \\
\good{12} & \good{161.28M} & \good{0.07} & \good{2.98} & \good{1.52} & \good{0.99} & \good{4.05} & \good{2.24} & \good{18.47} & \good{7.29} & \good{13.60} & \good{14.08} & \good{1.40} & \good{25.31} & \good{7.67}  & \good{18.40} \\
\good{13} & \good{176.88M} & \good{0.07} & \good{2.78} & \good{1.55} & \good{0.92} & \good{4.01} & \good{2.07} & \good{18.33} & \good{7.54} & \good{13.39} & \good{13.48} & \good{1.53} & \good{19.56} & \good{7.10}  & \good{14.61} \\
\good{14} & \good{189.48M} &\good{0.08} & \good{2.73} & \good{1.55} & \good{0.91} & \good{4.20} & \good{2.07} & \good{18.39} & \good{7.67} & \good{13.70} & \good{13.37} & \good{1.69} & \good{20.09} & \good{7.20}  & \good{15.07} \\
\good{15} & \good{202.07M} & \good{0.08} & \good{2.67} & \good{1.61} & \good{1.02} & \good{4.17} & \good{1.90} & \good{18.58} & \good{7.79} & \good{14.00} & \good{12.80} & \good{1.92} & \good{19.18} & \good{7.14}  & \good{14.41} \\
\good{16} & \good{214.67M} & \good{0.07} & \good{2.44} & \good{1.56} & \good{1.06} & \good{4.09} & \good{1.77} & \good{18.38} & \good{7.43} & \good{12.94} & \good{12.66} & \good{1.93} & \good{17.00} & \good{6.78}  & \good{12.87} \\
\good{17} & \good{227.26M} & \good{0.09} & \good{2.42} & \good{1.62} & \good{1.14} & \good{4.36} & \good{1.68} & \good{18.38} & \good{7.67} & \good{13.89} & \good{12.79} & \good{1.74} & \good{13.83} & \good{6.63}  & \good{12.48} \\
\good{18} & \good{239.86M} & \good{0.07} & \good{2.00} & \good{1.58} & \good{1.36} & \good{4.37} & \good{1.64} & \good{18.24} & \good{6.89} & \good{13.96} & \good{11.53} & \good{1.88} & \good{13.81} & \good{6.44}  & \good{12.55} \\
\good{19} & \good{252.46M} & \good{0.08} & \good{1.92} & \good{1.54} & \good{1.40} & \good{4.45} & \good{1.14} & \good{18.38} & \good{6.63} & \good{13.90} & \good{10.89} & \good{1.67} & \good{14.29} & \good{6.36}  & \good{13.42} \\
\good{20} & \good{265.05M} & \good{0.05} & \good{1.59} & \good{1.54} & \good{1.55} & \good{4.32} & \good{1.36} & \good{18.37} & \good{5.93} & \good{12.90} & \good{10.64} & \good{2.49} & \good{15.21} & \good{6.33}  & \good{13.81} \\
\good{21} & \good{277.65M} & \good{0.05} & \good{1.59} & \good{1.45} & \good{1.63} & \good{4.42} & \good{1.94} & \good{19.27} & \good{6.50} & \good{12.52} & \good{10.01} & \good{2.92} & \good{16.31} & \good{6.55}  & \good{13.86} \\
\good{22} & \good{290.25M} & \good{0.07} & \good{1.59} & \good{1.44} & \good{1.59} & \good{4.44} & \good{2.74} & \good{20.33} & \good{7.43} & \good{12.20} & \good{9.51}  & \good{2.62} & \good{16.77} & \good{6.73}  & \good{13.69} \\
\good{23} & \good{302.84M} & \good{0.08} & \good{1.71} & \good{1.49} & \good{1.59} & \good{3.98} & \good{2.44} & \good{18.60} & \good{9.29} & \good{12.28} & \good{9.57}  & \good{2.72} & \good{15.76} & \good{6.63}  & \good{13.21} \\
\good{24} & \good{315.44M} & \good{0.10} & \good{1.96} & \good{1.51} & \good{1.63} & \good{3.62} & \good{2.57} & \good{18.52} & \good{11.40} & \good{12.28} & \good{9.50}  & \good{3.83} & \good{14.97} & \good{6.82}  & \good{12.58} \\
\bottomrule
\multicolumn{16}{l}{\small Colour coding by Avg.\ EER:
  \colorbox{badcell}{red}~$>40\%$,\quad
  \colorbox{deadcell}{grey}~$12$--$40\%$,\quad
  \colorbox{goodcell}{green}~$<12\%$.}
\end{tabular}}
\end{table}

\begin{figure}[htbp]
    \centering

    \begin{minipage}{0.49\textwidth}
        \centering
        \includegraphics[width=\linewidth]{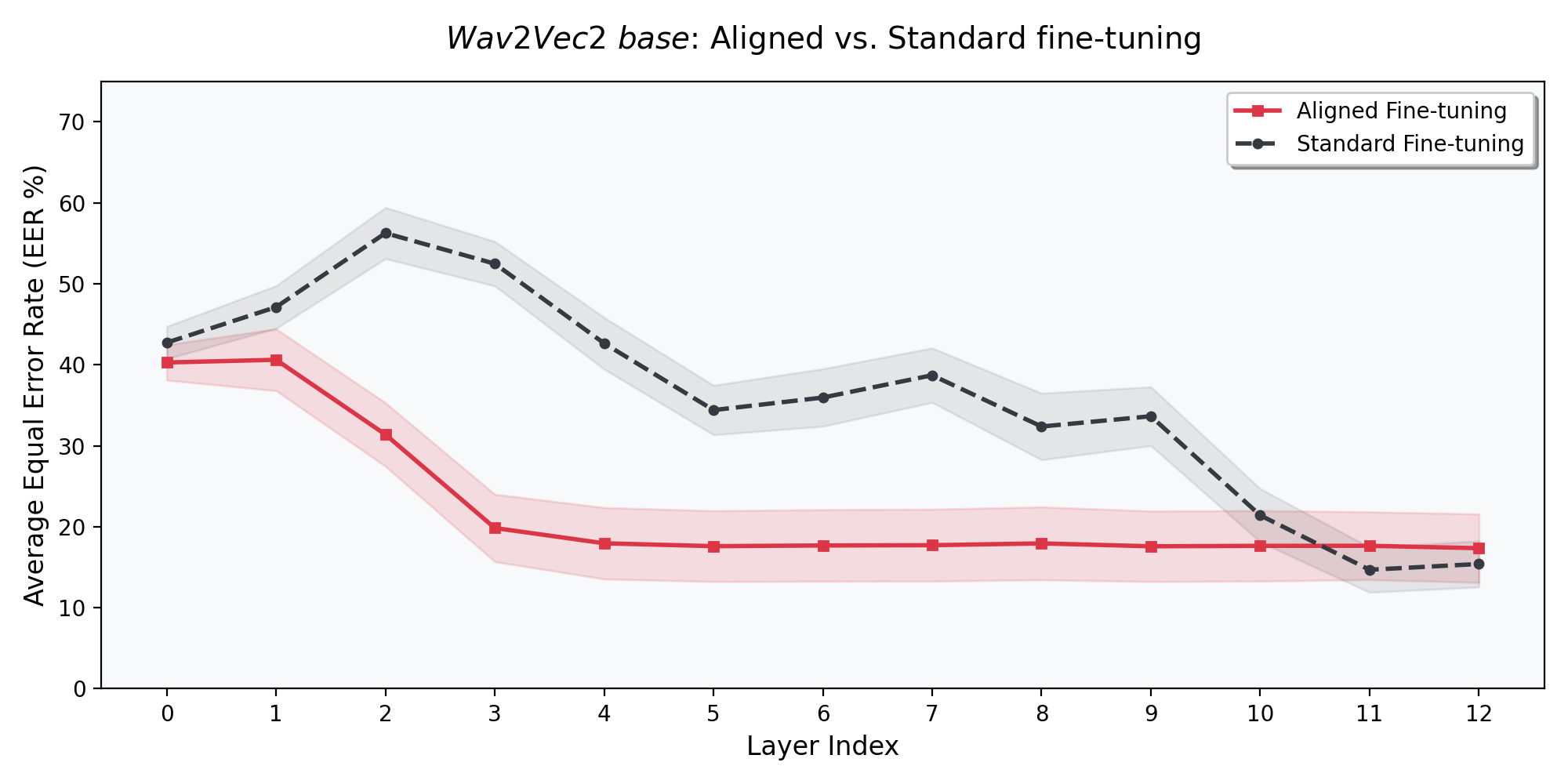}
    \end{minipage}\hfill
    \begin{minipage}{0.49\textwidth}
        \centering
        \includegraphics[width=\linewidth]{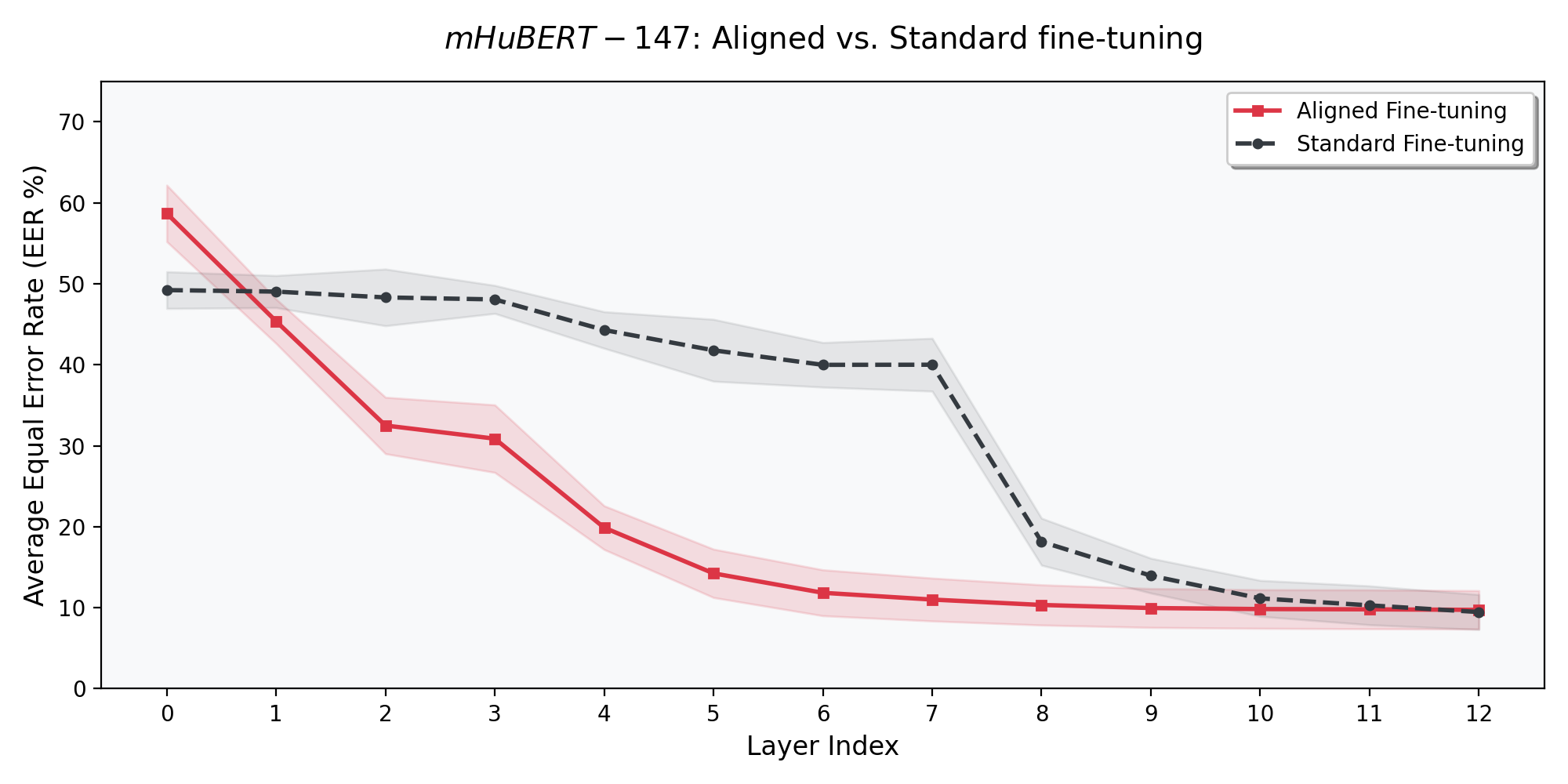}
    \end{minipage}\hfill

    \begin{minipage}{0.49\textwidth}
        \centering
        \includegraphics[width=\linewidth]{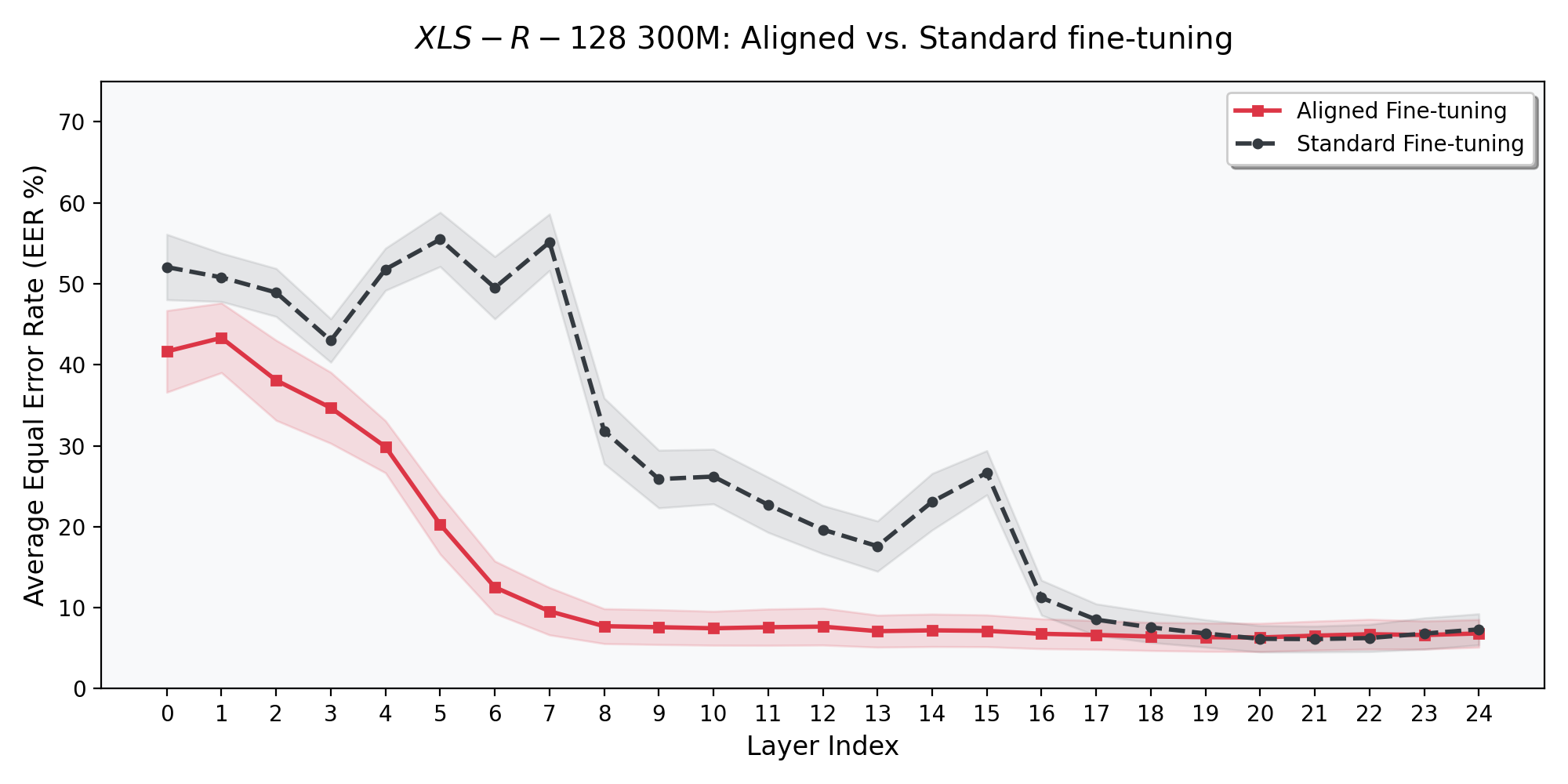}
    \end{minipage}\hfill
    \begin{minipage}{0.49\textwidth}
        \centering
        \includegraphics[width=\linewidth]{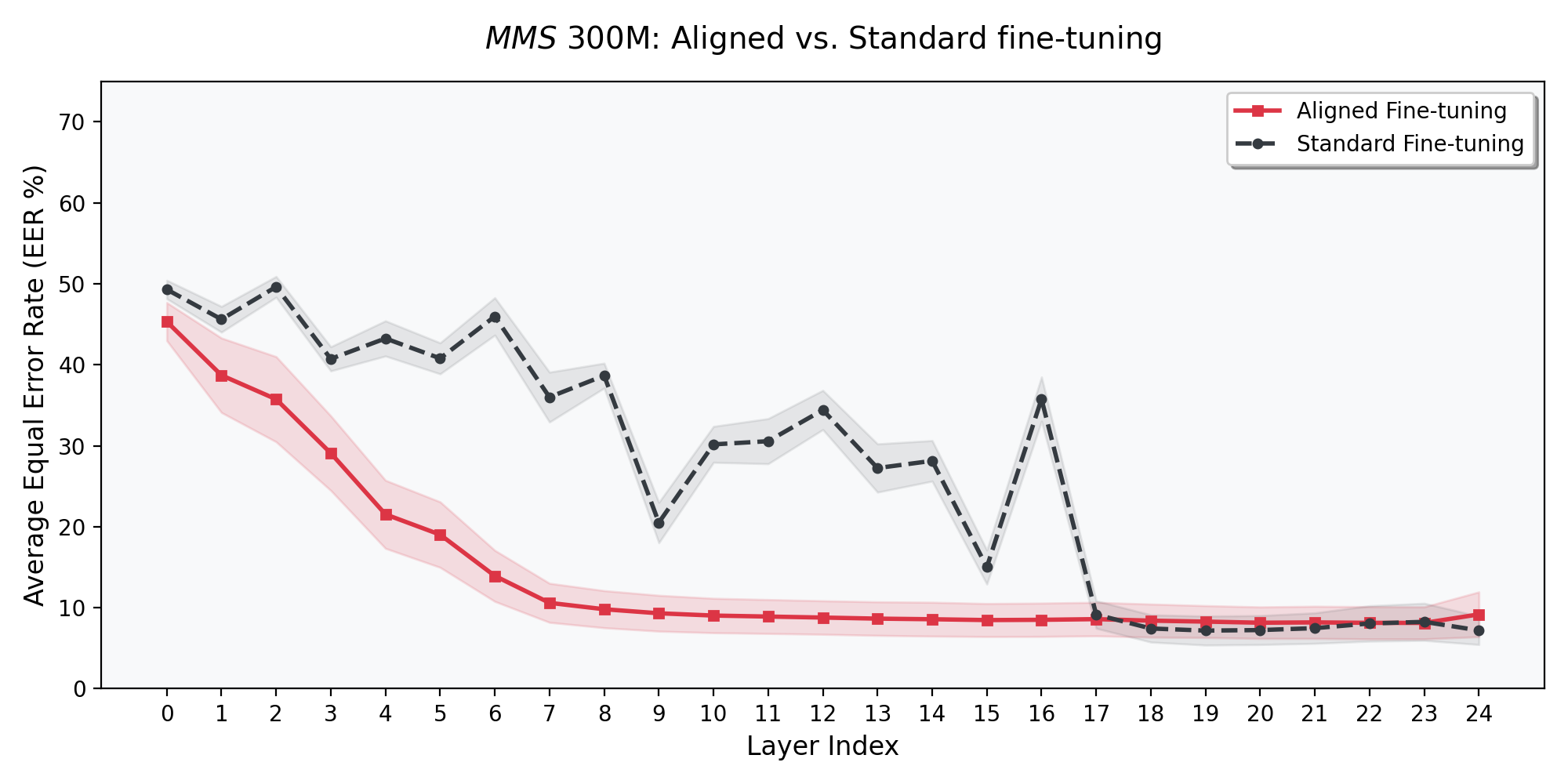}
    \end{minipage}\hfill

    \begin{minipage}{0.99\textwidth}
        \centering
        \includegraphics[width=0.8\linewidth]{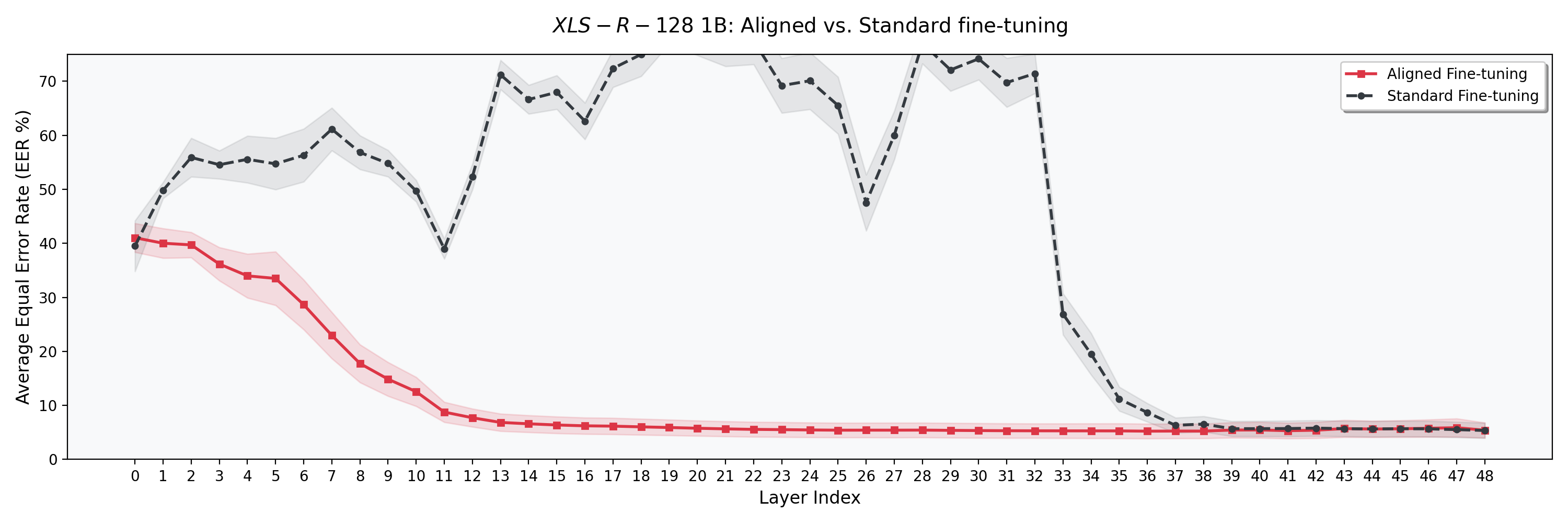}
    \end{minipage}\hfill

    \caption{Comparison (Average EER \%) of \emph{standard} and \emph{aligned} training across SFMs.
    }
    \label{fig:eer_comparison_all}
\end{figure}

\section{Results and analysis}
 \label{sec:results}
\subsection{Aligned training enabling early-exit}

\paragraph{Shallow-layer dormancy.}
Table~\ref{tab:xlsr300m_standard_aligned} confirms the dormancy problem for \emph{XLS-R-128} 300M under standard fine-tuning. Layers 0–7 yield average EER > 40\% — near-random performance. Competitive performance (EER < 12\%) emerges only from layer 16 onward, with best performance at layer 21 (6.13\% Avg. EER). The final layer (layer 24, 7.33\%) is outperformed by layer 21, suggesting mild overfitting to the training distribution at the topmost layer.

\paragraph{Aligned training compared with standard fine-tuning.}
Aligned training ($\rho=1$~\cite{jiang2025tracing}) shifts the layer-wise dynamics of \emph{XLS-R-128} 300M, significantly improving early and intermediate representation utility. Notably, aligned training achieves competitive performance by layer 7 (9.56\%), whereas standard fine-tuning requires 16 layers (11.23\%)—a 9-layer inference saving. It also stabilizes performance: while standard fine-tuning fluctuates (6.13–31.84\% across layers 8–24), aligned training remains robust (6.33–7.72\%), enabling early-exit.

\paragraph{Effect of model scale and pretraining language.}
Figure~\ref{fig:eer_comparison_all} highlights two consistent trends across architectures. First, scaling model capacity substantially improves robustness, especially for multilingual models. Under standard fine-tuning, the average EER of multilingual models decreases steadily from 9.46\% for \emph{mHuBERT-147} (base) to 5.80\% for \emph{XLS-R-128} (large) and further to 5.36\% for \emph{XLS-R-128} (XL), indicating that larger models produce more transferable representations for spoofing detection. Second, aligned fine-tuning improves layer efficiency by shifting the optimal decision layer earlier, while preserving competitive final-layer performance. For example, the best-performing layer of \emph{XLS-R-128 XL} moves from layer 48 under standard fine-tuning to layer 36 under aligned fine-tuning, while the average EER improves slightly from 5.36 to 5.23. This suggests that alignment encourages discriminative representations to emerge earlier in the network, reducing inference depth without sacrificing performance.

\begin{figure}[htbp]
  \centering
  \includegraphics[width=\textwidth]{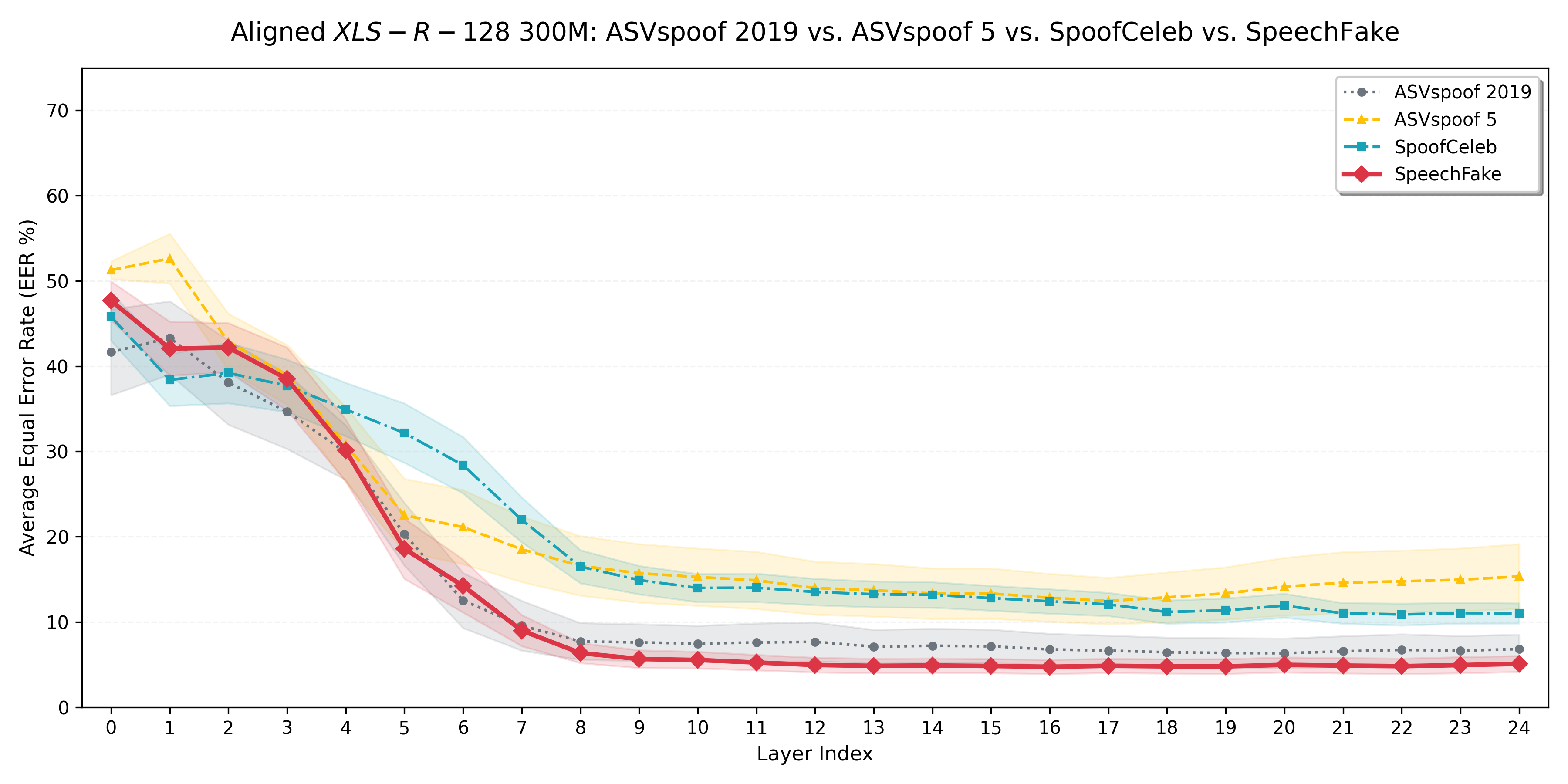}
  \caption{Performance (Average EER \%) of aligned fine-tuning trained on different datasets.}
  \label{fig:aligned_datasets}
\end{figure}
\subsection{Impact of training datasets on layer-wise generalization}
\label{sec:impact_training_data}

We next analyze how the choice of training dataset affects the layer-wise behavior of the model across diverse evaluation benchmarks. Figure~\ref{fig:aligned_datasets} reports the EER obtained by probing each encoder layer when the model is trained separately on four datasets with different characteristics: \texttt{ASVspoof 2019}, \texttt{ASVspoof 5}, \texttt{SpoofCeleb}, and \texttt{SpeechFake}. Detail results are described in Appendix~\ref{sec:impact_training_data_appendix}.

These results reveal that training data diversity and size of the training corpus has a larger effect on robustness.
Conversely, datasets such as \texttt{SpeechFake}, which contain broader acoustic and synthesis variability, enable the model to learn more invariant spoof-related cues, leading to substantially better cross-dataset generalization.
Overall, these results suggest that model scale and multilingual pretraining are the primary drivers of generalization, while aligned fine-tuning primarily improves computational efficiency by making strong representations emerge earlier. Based on its superior trade-off between robustness and early-layer efficiency, \emph{XLS-R-128} is utilized as our main backbone. Further analysis is conducted via PLS on the \texttt{19LA} training set, chosen specifically for its smaller scale in comparison to the more expansive \texttt{SpeechFake} corpus.

\subsection{Effect of PLS on earlier representation}

\paragraph{Effect of power parameter $\rho$.}
We ablate the layer-weight allocation in the aligned supervision objective (Eq.~\ref{eq:2}) using fixed exponents $\rho \in \{1, 1.5, 2\}$. This isolates the impact of weighting bias before introducing the dynamic schedule (Sec.~\ref{sec:method}). As shown in Table~\ref{tab:ablation_power}, $\rho$ governs the trade-off between depth specialization and distributed representation learning:

\begin{itemize}
\item \textbf{Low bias ($\rho=1$):} Provides relatively uniform supervision, allowing discriminative capacity to emerge early. Average EER falls below 12\% by layer 7 (11.11\%), though final-layer performance (7.36\%) lags behind higher exponents.
\item \textbf{Moderate bias ($\rho=1.5$):} Strikes the best balance. It optimizes intermediate layers smoothly (7.23\% EER at layer 9) while achieving the best overall Avg.\ EER (6.40\%) at layer 19. This suggests that a moderate deeper-layer preference guides high-level abstraction without starving shallow layers.
\item \textbf{High bias ($\rho=2$):} Concentrates optimization near the output, resembling standard fine-tuning. Early layers remain under-optimized (28.82\% EER at layer 5 vs.\ 22.33\% for $\rho=1.5$), effectively wasting the representational capacity of the intermediate backbone.
\end{itemize}

\begin{table}[htbp]
\centering
\caption{%
  Ablation of static power parameter $\rho$ (no exponential decay) on \emph{XLS-R-128} 300M
  with aligned fine-tuning. \textbf{Layer}: best-performing layer for each $\rho$.
  EER (\%, $\downarrow$).
}
\label{tab:ablation_power}
\resizebox{0.9\textwidth}{!}{%
\begin{tabular}{ccccccccccccccc}
\toprule
\multirow{2}{*}{$\boldsymbol{\rho}$} &
\multirow{2}{*}{\textbf{Layer}} &
\multicolumn{3}{c}{\textbf{In-Domain}} &
\textbf{Voc.} &
\multicolumn{3}{c}{\textbf{In-the-Wild}} &
\multicolumn{2}{c}{\textbf{Diff/Flow}} &
\multicolumn{3}{c}{\textbf{Other OOD}} &
\multirow{2}{*}{\textbf{Avg.}} \\
\cmidrule(lr){3-5}\cmidrule(lr){6-6}\cmidrule(lr){7-9}\cmidrule(lr){10-11}\cmidrule(lr){12-14}
& & \texttt{19LA} & \texttt{21LA} & \texttt{21DF} & \texttt{LSV}
& \texttt{ITW} & \texttt{FoR} & \texttt{SC}
& \texttt{DFADD} & \texttt{DIFSSD}
& \texttt{MLAAD-EN} & \texttt{DECRO-EN} & \texttt{25DF} & \\
\midrule

1 & 23 & \textbf{0.08} & \textbf{1.71} & {1.49} & 1.59 & \textbf{3.98} & 2.44 & 18.60 & 9.29 & 12.28 & \textbf{9.57} & 2.72 & 15.76 & 7.36 \\
1.5 & 19 & \underline{0.09} & 2.29 & \textbf{1.33} & \textbf{1.44} & \underline{4.88} & \underline{2.20} & \underline{17.83} & \textbf{7.41} & \textbf{10.26} & \underline{14.06} & \underline{1.79} & \underline{13.27} & \textbf{6.40} \\
2 & 19 & 0.10 & \underline{1.84} & \underline{1.43} & \underline{1.51} & 5.41 & \textbf{1.73} & \textbf{16.99} & \underline{8.21} & \underline{10.58} & 14.48 & \textbf{2.16} & \textbf{13.19} & \underline{6.47} \\

\bottomrule
\multicolumn{15}{l}{\normalsize \textbf{Bold} indicates best results, the second-best are \underline{underlined}.}
\end{tabular}}
\end{table}

These results indicate that while deeper-layer emphasis is necessary for peak performance, over-concentration hinders the "early-exit" utility of the model. This motivates a dynamic supervision that relaxes from a high $\rho$ toward a more uniform distribution.

\begin{figure}[htbp] 
  \centering
  \includegraphics[width=\linewidth]{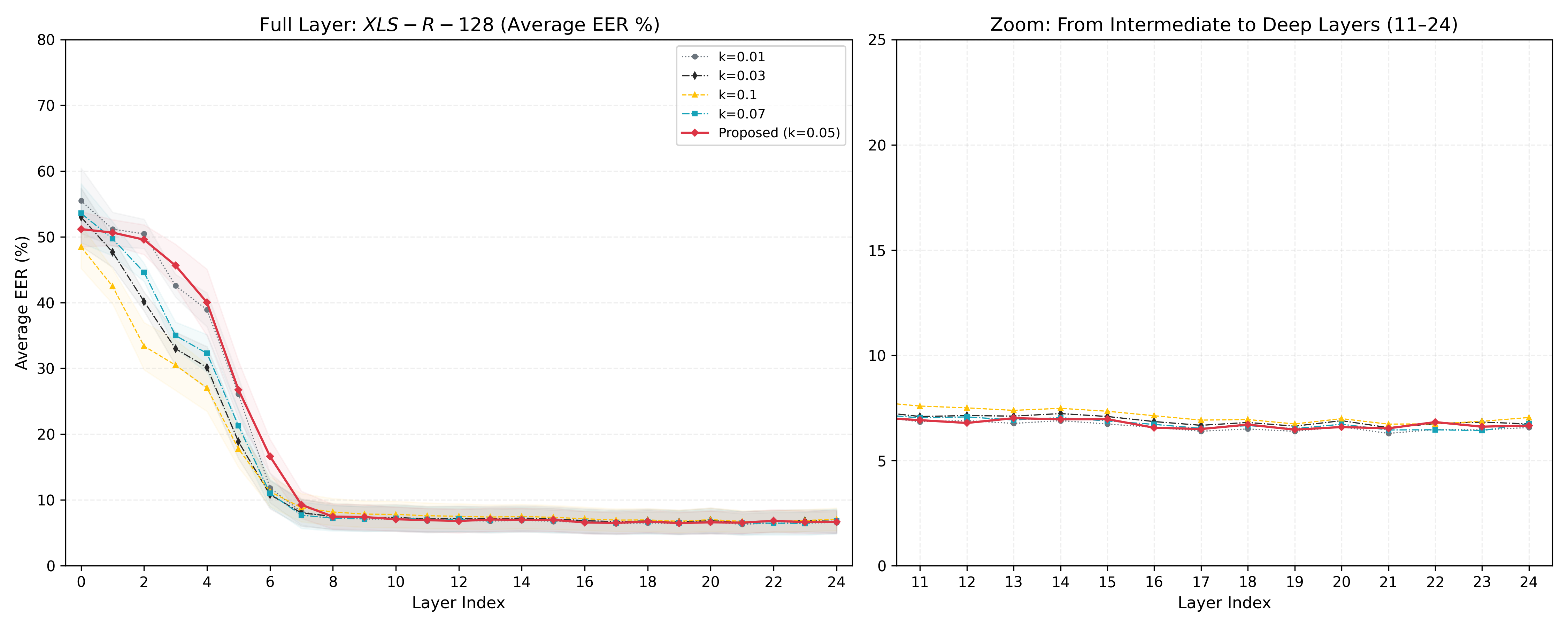}
  \caption{Effect of PLS with $M=1.5$ and $k$ decay.}
  \label{fig:power_k_layerwise}
\end{figure}

\paragraph{Effect of decay rate $k$.} We further evaluate the PLS dynamics by varying the decay rate $k \in \{0.01, 0.03, 0.05, 0.07, 0.1\}$ with fixed $M=1.5$ (Fig.~\ref{fig:power_k_layerwise}). The decay rate controls the transition from deep-layer initialization to holistic supervision. A slow decay ($k=0.01$) fixes the model's focus on deep layers too long, delaying the optimization of earlier features. Conversely, a rapid decay ($k=0.1$) collapses the supervision prematurely into uniform weighting, losing the stability benefits of the initial deep-layer bias. We find $k=0.05$ provides the optimal trade-off, allowing the model to stabilize high-level features before progressively refining the entire encoder for better generalization.

\begin{table}[htbp]
\centering
\caption{\emph{XLS-R-128} 300M under \emph{PLS} fine-tuning: EER (\%, $\downarrow$) at each transformer.}
\label{tab:xlsr_m15_k001_only}
\resizebox{\textwidth}{!}{%
\begin{tabular}{ccccccccccccccc}
\toprule
\multirow{2}{*}{\textbf{Layer}} & \multicolumn{3}{c}{\textbf{In-Domain}} & \textbf{Voc.} & \multicolumn{3}{c}{\textbf{In-the-Wild}} & \multicolumn{2}{c}{\textbf{Diff/Flow}} & \multicolumn{3}{c}{\textbf{Other OOD}} & \multicolumn{2}{c}{\textbf{Summary}} \\ \cmidrule(lr){2-4}\cmidrule(lr){5-5}\cmidrule(lr){6-8} \cmidrule(lr){9-10}\cmidrule(lr){11-13} \cmidrule(lr){14-15}
& 19LA & 21LA & 21DF & LSV & ITW & FoR & SC & DFADD & DIFSSD & MLAAD-EN & DECRO-EN & 25DF & \textbf{Avg.} & \textbf{Pool.} \\
\midrule
0 & \cellcolor[rgb]{0.78, 0.78, 0.78} 50.48 & \cellcolor[rgb]{0.79, 0.79, 0.79} 49.39 & \cellcolor[rgb]{0.78, 0.78, 0.78} 50.62 & \cellcolor[rgb]{0.78, 0.78, 0.78} 50.89 & \cellcolor[rgb]{0.84, 0.84, 0.84} 40.74 & \cellcolor[rgb]{0.78, 0.78, 0.78} 51.06 & \cellcolor[rgb]{0.81, 0.81, 0.81} 47.09 & \cellcolor[rgb]{0.73, 0.73, 0.73} 59.07 & \cellcolor[rgb]{0.86, 0.86, 0.86} 36.46 & \cellcolor[rgb]{0.77, 0.77, 0.77} 53.00 & \cellcolor[rgb]{0.60, 0.60, 0.60} 76.06 & \cellcolor[rgb]{0.80, 0.80, 0.80} 49.12 & \cellcolor[rgb]{0.78, 0.78, 0.78} 51.16 & \cellcolor[rgb]{0.81, 0.81, 0.81} 46.50\\
1 & \cellcolor[rgb]{0.76, 0.76, 0.76} 54.88 & \cellcolor[rgb]{0.72, 0.72, 0.72} 61.57 & \cellcolor[rgb]{0.80, 0.80, 0.80} 49.11 & \cellcolor[rgb]{0.81, 0.81, 0.81} 47.06 & \cellcolor[rgb]{0.82, 0.82, 0.82} 43.94 & \cellcolor[rgb]{0.77, 0.77, 0.77} 53.50 & \cellcolor[rgb]{0.84, 0.84, 0.84} 41.52 & \cellcolor[rgb]{0.76, 0.76, 0.76} 55.23 & \cellcolor[rgb]{0.82, 0.82, 0.82} 44.15 & \cellcolor[rgb]{0.79, 0.79, 0.79} 49.65 & \cellcolor[rgb]{0.70, 0.70, 0.70} 63.21 & \cellcolor[rgb]{0.82, 0.82, 0.82} 43.88 & \cellcolor[rgb]{0.78, 0.78, 0.78} 50.64 & \cellcolor[rgb]{0.80, 0.80, 0.80} 47.58\\
2 & \cellcolor[rgb]{0.79, 0.79, 0.79} 49.98 & \cellcolor[rgb]{0.74, 0.74, 0.74} 57.82 & \cellcolor[rgb]{0.82, 0.82, 0.82} 43.82 & \cellcolor[rgb]{0.78, 0.78, 0.78} 51.95 & \cellcolor[rgb]{0.83, 0.83, 0.83} 42.56 & \cellcolor[rgb]{0.81, 0.81, 0.81} 46.46 & \cellcolor[rgb]{0.82, 0.82, 0.82} 43.98 & \cellcolor[rgb]{0.73, 0.73, 0.73} 59.73 & \cellcolor[rgb]{0.80, 0.80, 0.80} 47.57 & \cellcolor[rgb]{0.68, 0.68, 0.68} 66.30 & \cellcolor[rgb]{0.86, 0.86, 0.86} 37.37 & \cellcolor[rgb]{0.81, 0.81, 0.81} 47.25 & \cellcolor[rgb]{0.79, 0.79, 0.79} 49.57 & \cellcolor[rgb]{0.78, 0.78, 0.78} 51.83\\
3 & \cellcolor[rgb]{0.87, 0.87, 0.87} 35.15 & \cellcolor[rgb]{0.79, 0.79, 0.79} 49.74 & \cellcolor[rgb]{0.87, 0.87, 0.87} 35.46 & \cellcolor[rgb]{0.77, 0.77, 0.77} 52.74 & \cellcolor[rgb]{0.77, 0.77, 0.77} 52.93 & \cellcolor[rgb]{0.65, 0.65, 0.65} 70.33 & \cellcolor[rgb]{0.82, 0.82, 0.82} 44.47 & \cellcolor[rgb]{0.83, 0.83, 0.83} 42.26 & \cellcolor[rgb]{0.86, 0.86, 0.86} 37.83 & \cellcolor[rgb]{0.76, 0.76, 0.76} 55.79 & \cellcolor[rgb]{0.91, 0.91, 0.91} 25.71 & \cellcolor[rgb]{0.82, 0.82, 0.82} 45.15 & \cellcolor[rgb]{0.81, 0.81, 0.81} 45.63 & \cellcolor[rgb]{0.81, 0.81, 0.81} 47.20\\
4 & \cellcolor[rgb]{0.93, 0.93, 0.93} 21.68 & \cellcolor[rgb]{0.90, 0.90, 0.90} 27.52 & \cellcolor[rgb]{0.90, 0.90, 0.90} 29.42 & \cellcolor[rgb]{0.78, 0.78, 0.78} 51.53 & \cellcolor[rgb]{0.72, 0.72, 0.72} 60.82 & \cellcolor[rgb]{0.59, 0.59, 0.59} 77.77 & \cellcolor[rgb]{0.76, 0.76, 0.76} 55.63 & \cellcolor[rgb]{0.92, 0.92, 0.92} 25.05 & \cellcolor[rgb]{0.84, 0.84, 0.84} 40.42 & \cellcolor[rgb]{0.92, 0.92, 0.92} 25.47 & \cellcolor[rgb]{0.94, 0.94, 0.94} 20.27 & \cellcolor[rgb]{0.82, 0.82, 0.82} 44.50 & \cellcolor[rgb]{0.85, 0.85, 0.85} 40.01 & \cellcolor[rgb]{0.84, 0.84, 0.84} 41.02\\
5 & \cellcolor[rgb]{0.98, 0.98, 0.98} 7.19 & \cellcolor[rgb]{0.97, 0.97, 0.97} 10.14 & \cellcolor[rgb]{0.96, 0.96, 0.96} 12.88 & \cellcolor[rgb]{0.90, 0.90, 0.90} 27.98 & \cellcolor[rgb]{0.85, 0.85, 0.85} 39.38 & \cellcolor[rgb]{0.72, 0.72, 0.72} 61.93 & \cellcolor[rgb]{0.87, 0.87, 0.87} 35.78 & \cellcolor[rgb]{0.93, 0.93, 0.93} 22.39 & \cellcolor[rgb]{0.84, 0.84, 0.84} 41.45 & \cellcolor[rgb]{0.92, 0.92, 0.92} 23.20 & \cellcolor[rgb]{0.96, 0.96, 0.96} 14.21 & \cellcolor[rgb]{0.92, 0.92, 0.92} 24.25 & \cellcolor[rgb]{0.91, 0.91, 0.91} 26.73 & \cellcolor[rgb]{0.90, 0.90, 0.90} 29.44\\
6 & \cellcolor[rgb]{0.99, 0.99, 0.99} 2.86 & \cellcolor[rgb]{0.98, 0.98, 0.98} 6.13 & \cellcolor[rgb]{0.98, 0.98, 0.98} 5.77 & \cellcolor[rgb]{0.96, 0.96, 0.96} 13.02 & \cellcolor[rgb]{0.95, 0.95, 0.95} 15.64 & \cellcolor[rgb]{0.88, 0.88, 0.88} 31.81 & \cellcolor[rgb]{0.90, 0.90, 0.90} 29.49 & \cellcolor[rgb]{0.95, 0.95, 0.95} 18.01 & \cellcolor[rgb]{0.93, 0.93, 0.93} 23.00 & \cellcolor[rgb]{0.92, 0.92, 0.92} 25.44 & \cellcolor[rgb]{0.97, 0.97, 0.97} 11.22 & \cellcolor[rgb]{0.95, 0.95, 0.95} 17.11 & \cellcolor[rgb]{0.95, 0.95, 0.95} 16.62 & \cellcolor[rgb]{0.94, 0.94, 0.94} 19.34\\
7 & \cellcolor[rgb]{1.00, 1.00, 1.00} 0.24 & \cellcolor[rgb]{0.99, 0.99, 0.99} 3.27 & \cellcolor[rgb]{0.99, 0.99, 0.99} 2.11 & \cellcolor[rgb]{0.99, 0.99, 0.99} 2.61 & \cellcolor[rgb]{0.98, 0.98, 0.98} 6.57 & \cellcolor[rgb]{0.97, 0.97, 0.97} 8.87 & \cellcolor[rgb]{0.93, 0.93, 0.93} 22.29 & \cellcolor[rgb]{0.97, 0.97, 0.97} 9.93 & \cellcolor[rgb]{0.96, 0.96, 0.96} 14.54 & \cellcolor[rgb]{0.94, 0.94, 0.94} 20.99 & \cellcolor[rgb]{0.99, 0.99, 0.99} 4.43 & \cellcolor[rgb]{0.96, 0.96, 0.96} 15.09 & \cellcolor[rgb]{0.97, 0.97, 0.97} 9.25 & \cellcolor[rgb]{0.96, 0.96, 0.96} 15.14\\
8 & \cellcolor[rgb]{1.00, 1.00, 1.00} 0.16 & \cellcolor[rgb]{0.99, 0.99, 0.99} 3.13 & \cellcolor[rgb]{1.00, 1.00, 1.00} 1.40 & \cellcolor[rgb]{1.00, 1.00, 1.00} 1.33 & \cellcolor[rgb]{0.99, 0.99, 0.99} 4.76 & \cellcolor[rgb]{0.99, 0.99, 0.99} 3.71 & \cellcolor[rgb]{0.94, 0.94, 0.94} 19.53 & \cellcolor[rgb]{0.97, 0.97, 0.97} 8.74 & \cellcolor[rgb]{0.97, 0.97, 0.97} 11.35 & \cellcolor[rgb]{0.95, 0.95, 0.95} 18.21 & \cellcolor[rgb]{0.99, 0.99, 0.99} 2.99 & \cellcolor[rgb]{0.96, 0.96, 0.96} 14.08 & \cellcolor[rgb]{0.98, 0.98, 0.98} 7.45 & \cellcolor[rgb]{0.96, 0.96, 0.96} 13.48\\
9 & \cellcolor[rgb]{1.00, 1.00, 1.00} 0.14 & \cellcolor[rgb]{0.99, 0.99, 0.99} 3.36 & \cellcolor[rgb]{1.00, 1.00, 1.00} 1.44 & \cellcolor[rgb]{1.00, 1.00, 1.00} 1.25 & \cellcolor[rgb]{0.99, 0.99, 0.99} 4.59 & \cellcolor[rgb]{0.99, 0.99, 0.99} 3.09 & \cellcolor[rgb]{0.94, 0.94, 0.94} 18.86 & \cellcolor[rgb]{0.97, 0.97, 0.97} 9.27 & \cellcolor[rgb]{0.96, 0.96, 0.96} 12.37 & \cellcolor[rgb]{0.95, 0.95, 0.95} 18.24 & \cellcolor[rgb]{0.99, 0.99, 0.99} 2.20 & \cellcolor[rgb]{0.96, 0.96, 0.96} 13.82 & \cellcolor[rgb]{0.98, 0.98, 0.98} 7.39 & \cellcolor[rgb]{0.96, 0.96, 0.96} 13.07\\
10 & \cellcolor[rgb]{1.00, 1.00, 1.00} 0.10 & \cellcolor[rgb]{0.99, 0.99, 0.99} 3.14 & \cellcolor[rgb]{1.00, 1.00, 1.00} 1.35 & \cellcolor[rgb]{1.00, 1.00, 1.00} 1.17 & \cellcolor[rgb]{0.99, 0.99, 0.99} 4.41 & \cellcolor[rgb]{0.99, 0.99, 0.99} 2.44 & \cellcolor[rgb]{0.94, 0.94, 0.94} 18.60 & \cellcolor[rgb]{0.98, 0.98, 0.98} 8.46 & \cellcolor[rgb]{0.96, 0.96, 0.96} 11.71 & \cellcolor[rgb]{0.95, 0.95, 0.95} 17.19 & \cellcolor[rgb]{0.99, 0.99, 0.99} 2.28 & \cellcolor[rgb]{0.96, 0.96, 0.96} 13.69 & \cellcolor[rgb]{0.98, 0.98, 0.98} 7.04 & \cellcolor[rgb]{0.96, 0.96, 0.96} 12.84\\
11 & \cellcolor[rgb]{1.00, 1.00, 1.00} 0.11 & \cellcolor[rgb]{0.99, 0.99, 0.99} 2.96 & \cellcolor[rgb]{1.00, 1.00, 1.00} 1.32 & \cellcolor[rgb]{1.00, 1.00, 1.00} 1.06 & \cellcolor[rgb]{0.99, 0.99, 0.99} 4.35 & \cellcolor[rgb]{0.99, 0.99, 0.99} 2.16 & \cellcolor[rgb]{0.95, 0.95, 0.95} 18.26 & \cellcolor[rgb]{0.97, 0.97, 0.97} 8.87 & \cellcolor[rgb]{0.97, 0.97, 0.97} 11.10 & \cellcolor[rgb]{0.95, 0.95, 0.95} 17.10 & \cellcolor[rgb]{0.99, 0.99, 0.99} 2.32 & \cellcolor[rgb]{0.96, 0.96, 0.96} 13.32 & \cellcolor[rgb]{0.98, 0.98, 0.98} 6.91 & \cellcolor[rgb]{0.96, 0.96, 0.96} 12.64\\
12 & \cellcolor[rgb]{1.00, 1.00, 1.00} 0.10 & \cellcolor[rgb]{0.99, 0.99, 0.99} 3.25 & \cellcolor[rgb]{1.00, 1.00, 1.00} 1.37 & \cellcolor[rgb]{1.00, 1.00, 1.00} 1.10 & \cellcolor[rgb]{0.99, 0.99, 0.99} 4.37 & \cellcolor[rgb]{1.00, 1.00, 1.00} 1.81 & \cellcolor[rgb]{0.95, 0.95, 0.95} 18.00 & \cellcolor[rgb]{0.97, 0.97, 0.97} 8.87 & \cellcolor[rgb]{0.97, 0.97, 0.97} 10.44 & \cellcolor[rgb]{0.95, 0.95, 0.95} 16.85 & \cellcolor[rgb]{1.00, 1.00, 1.00} 1.88 & \cellcolor[rgb]{0.96, 0.96, 0.96} 13.29 & \cellcolor[rgb]{0.98, 0.98, 0.98} 6.78 & \cellcolor[rgb]{0.96, 0.96, 0.96} 12.65\\
13 & \cellcolor[rgb]{1.00, 1.00, 1.00} 0.10 & \cellcolor[rgb]{0.99, 0.99, 0.99} 2.93 & \cellcolor[rgb]{1.00, 1.00, 1.00} 1.39 & \cellcolor[rgb]{1.00, 1.00, 1.00} 1.10 & \cellcolor[rgb]{0.99, 0.99, 0.99} 4.44 & \cellcolor[rgb]{0.99, 0.99, 0.99} 2.29 & \cellcolor[rgb]{0.95, 0.95, 0.95} 17.94 & \cellcolor[rgb]{0.97, 0.97, 0.97} 9.93 & \cellcolor[rgb]{0.97, 0.97, 0.97} 11.32 & \cellcolor[rgb]{0.95, 0.95, 0.95} 16.94 & \cellcolor[rgb]{0.99, 0.99, 0.99} 2.39 & \cellcolor[rgb]{0.96, 0.96, 0.96} 13.30 & \cellcolor[rgb]{0.98, 0.98, 0.98} 7.01 & \cellcolor[rgb]{0.96, 0.96, 0.96} 12.57\\
14 & \cellcolor[rgb]{1.00, 1.00, 1.00} 0.11 & \cellcolor[rgb]{0.99, 0.99, 0.99} 2.83 & \cellcolor[rgb]{1.00, 1.00, 1.00} 1.37 & \cellcolor[rgb]{1.00, 1.00, 1.00} 1.14 & \cellcolor[rgb]{0.99, 0.99, 0.99} 4.32 & \cellcolor[rgb]{0.99, 0.99, 0.99} 2.61 & \cellcolor[rgb]{0.95, 0.95, 0.95} 17.91 & \cellcolor[rgb]{0.97, 0.97, 0.97} 10.08 & \cellcolor[rgb]{0.97, 0.97, 0.97} 11.25 & \cellcolor[rgb]{0.95, 0.95, 0.95} 16.50 & \cellcolor[rgb]{0.99, 0.99, 0.99} 2.09 & \cellcolor[rgb]{0.96, 0.96, 0.96} 13.24 & \cellcolor[rgb]{0.98, 0.98, 0.98} 6.95 & \cellcolor[rgb]{0.96, 0.96, 0.96} 12.53\\
15 & \cellcolor[rgb]{1.00, 1.00, 1.00} 0.08 & \cellcolor[rgb]{0.99, 0.99, 0.99} 2.50 & \cellcolor[rgb]{1.00, 1.00, 1.00} 1.35 & \cellcolor[rgb]{1.00, 1.00, 1.00} 1.14 & \cellcolor[rgb]{0.99, 0.99, 0.99} 4.27 & \cellcolor[rgb]{0.99, 0.99, 0.99} 3.00 & \cellcolor[rgb]{0.95, 0.95, 0.95} 17.94 & \cellcolor[rgb]{0.97, 0.97, 0.97} 9.93 & \cellcolor[rgb]{0.96, 0.96, 0.96} 11.81 & \cellcolor[rgb]{0.95, 0.95, 0.95} 15.71 & \cellcolor[rgb]{0.99, 0.99, 0.99} 2.51 & \cellcolor[rgb]{0.96, 0.96, 0.96} 13.26 & \cellcolor[rgb]{0.98, 0.98, 0.98} 6.96 & \cellcolor[rgb]{0.96, 0.96, 0.96} 12.56\\
16 & \cellcolor[rgb]{1.00, 1.00, 1.00} 0.08 & \cellcolor[rgb]{0.99, 0.99, 0.99} 2.46 & \cellcolor[rgb]{1.00, 1.00, 1.00} 1.30 & \cellcolor[rgb]{1.00, 1.00, 1.00} 1.21 & \cellcolor[rgb]{0.99, 0.99, 0.99} 4.15 & \cellcolor[rgb]{0.99, 0.99, 0.99} 2.29 & \cellcolor[rgb]{0.95, 0.95, 0.95} 17.75 & \cellcolor[rgb]{0.97, 0.97, 0.97} 8.90 & \cellcolor[rgb]{0.97, 0.97, 0.97} 10.49 & \cellcolor[rgb]{0.96, 0.96, 0.96} 14.78 & \cellcolor[rgb]{0.99, 0.99, 0.99} 2.15 & \cellcolor[rgb]{0.96, 0.96, 0.96} 13.07 & \cellcolor[rgb]{0.98, 0.98, 0.98} 6.55 & \cellcolor[rgb]{0.96, 0.96, 0.96} 12.41\\
17 & \cellcolor[rgb]{1.00, 1.00, 1.00} 0.10 & \cellcolor[rgb]{0.99, 0.99, 0.99} 2.65 & \cellcolor[rgb]{1.00, 1.00, 1.00} 1.28 & \cellcolor[rgb]{1.00, 1.00, 1.00} 1.18 & \cellcolor[rgb]{0.99, 0.99, 0.99} 4.16 & \cellcolor[rgb]{0.99, 0.99, 0.99} 2.07 & \cellcolor[rgb]{0.95, 0.95, 0.95} 17.68 & \cellcolor[rgb]{0.98, 0.98, 0.98} 8.22 & \cellcolor[rgb]{0.97, 0.97, 0.97} 10.21 & \cellcolor[rgb]{0.95, 0.95, 0.95} 15.27 & \cellcolor[rgb]{0.99, 0.99, 0.99} 2.06 & \cellcolor[rgb]{0.96, 0.96, 0.96} 12.99 & \cellcolor[rgb]{0.98, 0.98, 0.98} 6.49 & \cellcolor[rgb]{0.96, 0.96, 0.96} 12.25\\
18 & \cellcolor[rgb]{1.00, 1.00, 1.00} 0.08 & \cellcolor[rgb]{0.99, 0.99, 0.99} 2.58 & \cellcolor[rgb]{1.00, 1.00, 1.00} 1.33 & \cellcolor[rgb]{1.00, 1.00, 1.00} 1.21 & \cellcolor[rgb]{0.99, 0.99, 0.99} 4.27 & \cellcolor[rgb]{0.99, 0.99, 0.99} 2.29 & \cellcolor[rgb]{0.95, 0.95, 0.95} 17.93 & \cellcolor[rgb]{0.97, 0.97, 0.97} 9.17 & \cellcolor[rgb]{0.97, 0.97, 0.97} 11.07 & \cellcolor[rgb]{0.96, 0.96, 0.96} 15.14 & \cellcolor[rgb]{0.99, 0.99, 0.99} 2.16 & \cellcolor[rgb]{0.96, 0.96, 0.96} 13.18 & \cellcolor[rgb]{0.98, 0.98, 0.98} 6.70 & \cellcolor[rgb]{0.96, 0.96, 0.96} 12.30\\
19 & \cellcolor[rgb]{1.00, 1.00, 1.00} 0.10 & \cellcolor[rgb]{0.99, 0.99, 0.99} 2.74 & \cellcolor[rgb]{1.00, 1.00, 1.00} 1.35 & \cellcolor[rgb]{1.00, 1.00, 1.00} 1.21 & \cellcolor[rgb]{0.99, 0.99, 0.99} 4.13 & \cellcolor[rgb]{0.99, 0.99, 0.99} 2.61 & \cellcolor[rgb]{0.94, 0.94, 0.94} 18.35 & \cellcolor[rgb]{0.98, 0.98, 0.98} 7.97 & \cellcolor[rgb]{0.97, 0.97, 0.97} 10.39 & \cellcolor[rgb]{0.96, 0.96, 0.96} 14.06 & \cellcolor[rgb]{1.00, 1.00, 1.00} 1.70 & \cellcolor[rgb]{0.96, 0.96, 0.96} 12.91 & \cellcolor[rgb]{0.98, 0.98, 0.98} 6.46 & \cellcolor[rgb]{0.96, 0.96, 0.96} 12.20\\
20 & \cellcolor[rgb]{1.00, 1.00, 1.00} 0.11 & \cellcolor[rgb]{0.99, 0.99, 0.99} 3.16 & \cellcolor[rgb]{1.00, 1.00, 1.00} 1.35 & \cellcolor[rgb]{1.00, 1.00, 1.00} 1.11 & \cellcolor[rgb]{0.99, 0.99, 0.99} 4.01 & \cellcolor[rgb]{0.99, 0.99, 0.99} 2.65 & \cellcolor[rgb]{0.95, 0.95, 0.95} 17.93 & \cellcolor[rgb]{0.98, 0.98, 0.98} 8.21 & \cellcolor[rgb]{0.97, 0.97, 0.97} 10.58 & \cellcolor[rgb]{0.95, 0.95, 0.95} 15.74 & \cellcolor[rgb]{1.00, 1.00, 1.00} 1.09 & \cellcolor[rgb]{0.96, 0.96, 0.96} 13.07 & \cellcolor[rgb]{0.98, 0.98, 0.98} 6.58 & \cellcolor[rgb]{0.96, 0.96, 0.96} 12.22\\
21 & \cellcolor[rgb]{1.00, 1.00, 1.00} 0.10 & \cellcolor[rgb]{0.99, 0.99, 0.99} 2.67 & \cellcolor[rgb]{1.00, 1.00, 1.00} 1.30 & \cellcolor[rgb]{1.00, 1.00, 1.00} 1.18 & \cellcolor[rgb]{0.99, 0.99, 0.99} 4.55 & \cellcolor[rgb]{0.99, 0.99, 0.99} 3.45 & \cellcolor[rgb]{0.94, 0.94, 0.94} 18.35 & \cellcolor[rgb]{0.98, 0.98, 0.98} 8.51 & \cellcolor[rgb]{0.97, 0.97, 0.97} 9.68 & \cellcolor[rgb]{0.96, 0.96, 0.96} 13.31 & \cellcolor[rgb]{1.00, 1.00, 1.00} 1.70 & \cellcolor[rgb]{0.96, 0.96, 0.96} 13.47 & \cellcolor[rgb]{0.98, 0.98, 0.98} 6.52 & \cellcolor[rgb]{0.96, 0.96, 0.96} 12.41\\
22 & \cellcolor[rgb]{1.00, 1.00, 1.00} 0.07 & \cellcolor[rgb]{0.99, 0.99, 0.99} 2.24 & \cellcolor[rgb]{1.00, 1.00, 1.00} 1.21 & \cellcolor[rgb]{1.00, 1.00, 1.00} 1.33 & \cellcolor[rgb]{0.99, 0.99, 0.99} 5.48 & \cellcolor[rgb]{0.99, 0.99, 0.99} 4.99 & \cellcolor[rgb]{0.94, 0.94, 0.94} 19.38 & \cellcolor[rgb]{0.98, 0.98, 0.98} 8.47 & \cellcolor[rgb]{0.97, 0.97, 0.97} 11.09 & \cellcolor[rgb]{0.97, 0.97, 0.97} 10.72 & \cellcolor[rgb]{0.99, 0.99, 0.99} 2.11 & \cellcolor[rgb]{0.96, 0.96, 0.96} 14.74 & \cellcolor[rgb]{0.98, 0.98, 0.98} 6.82 & \cellcolor[rgb]{0.96, 0.96, 0.96} 12.99\\
23 & \cellcolor[rgb]{1.00, 1.00, 1.00} 0.10 & \cellcolor[rgb]{0.99, 0.99, 0.99} 2.68 & \cellcolor[rgb]{1.00, 1.00, 1.00} 1.32 & \cellcolor[rgb]{1.00, 1.00, 1.00} 1.33 & \cellcolor[rgb]{0.99, 0.99, 0.99} 4.88 & \cellcolor[rgb]{0.99, 0.99, 0.99} 3.22 & \cellcolor[rgb]{0.94, 0.94, 0.94} 18.52 & \cellcolor[rgb]{0.98, 0.98, 0.98} 8.47 & \cellcolor[rgb]{0.97, 0.97, 0.97} 11.21 & \cellcolor[rgb]{0.97, 0.97, 0.97} 11.18 & \cellcolor[rgb]{1.00, 1.00, 1.00} 1.44 & \cellcolor[rgb]{0.96, 0.96, 0.96} 14.87 & \cellcolor[rgb]{0.98, 0.98, 0.98} 6.60 & \cellcolor[rgb]{0.96, 0.96, 0.96} 12.68\\
24 & \cellcolor[rgb]{1.00, 1.00, 1.00} 0.11 & \cellcolor[rgb]{0.99, 0.99, 0.99} 2.88 & \cellcolor[rgb]{1.00, 1.00, 1.00} 1.39 & \cellcolor[rgb]{1.00, 1.00, 1.00} 1.29 & \cellcolor[rgb]{0.99, 0.99, 0.99} 4.20 & \cellcolor[rgb]{0.99, 0.99, 0.99} 3.04 & \cellcolor[rgb]{0.94, 0.94, 0.94} 19.73 & \cellcolor[rgb]{0.98, 0.98, 0.98} 7.96 & \cellcolor[rgb]{0.97, 0.97, 0.97} 10.74 & \cellcolor[rgb]{0.97, 0.97, 0.97} 11.07 & \cellcolor[rgb]{0.99, 0.99, 0.99} 2.13 & \cellcolor[rgb]{0.95, 0.95, 0.95} 15.31 & \cellcolor[rgb]{0.98, 0.98, 0.98} 6.65 & \cellcolor[rgb]{0.96, 0.96, 0.96} 12.03\\

\bottomrule
\end{tabular}}
\end{table}

The results clearly highlight the advantage of the proposed PLS (Table~\ref{tab:xlsr_m15_k001_only}) over both standard and aligned fine-tuning (Table~\ref{tab:xlsr300m_standard_aligned}), particularly when considering the behavior of earlier layers. Unlike other approaches, PLS introduces a gradual and controlled supervision mechanism that better leverages the representational hierarchy of the model. This progressive strategy allows early layers to contribute meaningfully without being prematurely optimized, leading to a much more stable and consistent performance across diverse conditions. As a result, PLS avoids the large performance variance observed in the first layers of both baselines and achieves significantly improved generalization, especially on OOD and challenging scenarios. Moreover, while aligned fine-tuning tends to delay useful learning signals to deeper layers, PLS ensures a smoother transition across layers, yielding competitive—and often superior—performance not only in early layers but also maintaining strong results throughout the entire network. This demonstrates that PLS is not merely a training heuristic, but an effective paradigm for exploiting layer-wise learning dynamics in SFMs.

\begin{table}[htbp]
    \centering
    \caption{Overall performance (EER \%, $\downarrow$) compared to SOTA approaches. \textbf{Bold} indicates best results, the second-best are \underline{underlined}. $\dagger$ denotes results reported from~\cite{11345101}. $\diamond$ denotes results computed by using official released checkpoints.}
    \label{tab:sota_comparision}
    \resizebox{\linewidth}{!}{
    \begin{tabular}{lrccccccccc}
    \toprule
    \multirow{2}{*}{\textbf{Model}} & \multirow{2}{*}{\textbf{\begin{tabular}[c]{@{}c@{}}\#Params.\\(Inference)\end{tabular}}} & \multicolumn{3}{c}{\textbf{In-domain}} & \multicolumn{5}{c}{\textbf{Out-of-domain}} & \multirow{2}{*}{\textbf{Average}}\\ \cmidrule(lr){3-5} \cmidrule(lr){6-10}
    & & \texttt{\textbf{19LA}} & \texttt{\textbf{21LA}} & \texttt{\textbf{21DF}} & \texttt{\textbf{25DF}} & \texttt{\textbf{ITW}} & \texttt{\textbf{DFADD}} & \texttt{\textbf{LSV}} & \texttt{\textbf{SC}} \\ \midrule
    SLS\cite{zhang2024audio}$\dagger\diamond$ & 340M & 0.23 & 2.87 & 1.91 & 18.76 & 7.46 & 7.54 & 1.97 & 24.51 & 8.16 \\
    Conformer\cite{truong24b_interspeech}$\dagger\diamond$ & 319M & 0.19 & 1.03 & 2.06 & 18.85 & 7.79 & 8.89 & 2.35 & 38.15 & 9.91 \\
    MultiConv\cite{10.1145/3746027.3754568}$\diamond$ & 317M & \textbf{0.08} & 2.77 & 1.43 & 15.19 & 4.44 & \textbf{6.60} & 1.70 & 18.71 & 6.37 \\
    KAN\cite{11226534}$\diamond$ & 317M & 0.11 & 2.29 & 1.49 & 14.10 & 5.31 & 15.63 & 1.52 & 22.06 & 7.81 \\
    Mamba\cite{tran25b_interspeech}$\diamond$ & 317M & 0.11 & 1.78 & 1.51 & 13.58 & 5.12 & 8.62 & 1.82 & \underline{17.87} & 6.30 \\
    Nes2NetX\cite{11222612}$\dagger\diamond$ & 317M & 0.12 & 1.88 & 1.49 & 22.06 & 5.52 & 11.15 & 2.88 & 54.28 & 12.42\\
    AASIST\cite{tak22_odyssey}$\dagger\diamond$ & 317M & 0.22 & \textbf{0.82} & 2.85 & 16.25 & 11.20 & 11.93 & 11.21 & 26.72 & 10.15\\
    \midrule
    Baseline - FT & 315M & 0.10 & 2.36 & 1.46 & 13.47 & 4.69 & 17.51 & 1.75 & 19.57 & 7.61 \\
    Layer 17 (Ours) & 239M & 0.10 & 2.65 & \textbf{1.28} & \textbf{12.99} & \textbf{4.16} & 8.22 & \underline{1.18} & \textbf{17.68} & \textbf{6.03} \\ 
    Layer 11 (Ours) & 152M & 0.11 & 2.96 & \underline{1.32} & \underline{13.32} & \underline{4.35} & 8.87 & \textbf{1.06} & 18.26 & \underline{6.28} \\ 
    \bottomrule
    \end{tabular}%
    }
\end{table}

\paragraph{Detection performance versus SOTA.}

Table~\ref{tab:sota_comparision} reports the EER of the compared methods on both in-domain and OOD evaluation sets. Our proposed PLS models consistently improve over the standard fine-tuning baseline, especially in OOD conditions. 
For instance, \textit{Aligned - Layer 17} reduces the average EER from 7.61\% to 6.03\%, corresponding to a relative improvement of 20.8\%, while also outperforming the baseline on all OOD datasets. 
Notably, it achieves lower EERs on \texttt{25DF} (12.99\%), \texttt{ITW} (4.16\%), \texttt{DFADD} (8.22\%), \texttt{LSV} (1.18\%), and \texttt{SC} (17.68\%), indicating stronger robustness to unseen spoofing conditions. Compared with prior SOTA methods, \textit{PLS - Layer 17} achieves the best overall average EER of 6.03\%, outperforming \textit{Mamba} while using fewer trainable parameters (239M vs. 317M). 
Moreover, deeper pruning levels preserve strong detection capability: \textit{PLS - Layer 11} attains a competitive 6.28\% average EER with only 152M parameters, showing that the proposed alignment strategy enables efficient model compression without sacrificing generalization. 
These results suggest that the proposed PLS alignment improves the transferability of spoof-discriminative representations, leading to better cross-domain robustness with reduced model complexity.

\section{Discussion and limitation}
\label{sec:limitation}
We attribute OOD gains to increased representation diversity by distributing discriminative power across layers, PLS prevents the final-layer vulnerability to distribution shifts seen in standard models. This approach yields a 70\% reduction in inference parameters (101M vs. 315M) without the architectural overhead or teacher-dependency of distillation. However, these benefits are most evident in large multilingual models, as smaller backbones may lack the capacity to absorb intermediate supervision without performance trade-offs. Furthermore, while PLS optimizes inference, it still requires training the full model to convergence prior to pruning.

\section{Conclusion}
We introduced Progressive Layer-wise Supervision (PLS) to resolve shallow-layer dormancy in fine-tuned speech foundation models (SFMs). By utilizing exponentially annealed power weighting and shared classification, PLS activates early layers, enabling reliable early-exit inference nine layers sooner than standard method and achieving state-of-the-art out-of-domain generalization (6.03\% EER) comparing to prior approaches with fewer inference parameters. While PLS significantly reduces inference parameters, future work will focus on decreasing training time by analyzing relevant SFM hidden features to prune redundant parameters during the training phase itself.

\section*{Acknowledgements}
This work was granted access to the HPC/AI resources of IDRIS under the allocation made by GENCI, as well as the GPU resources provided by the Expression team at IRISA. This work was partially supported by JST PRESTO, Japan (Grant Number JPMJPR23P9).

\bibliographystyle{unsrtnat} 
\bibliography{references}

\newpage
\newpage

\section*{NeurIPS Paper Checklist}

\begin{enumerate}

\item {\bf Claims}
    \item[] Question: Do the main claims made in the abstract and introduction accurately reflect the paper's contributions and scope?
    \item[] Answer: \answerYes{} 
    \item[] Justification: The contributions and scope of our study are listed at the end of Section~\ref{sec:introduction}, and summarized in the abstract. The experimental results in Section~\ref{sec:results} support our claims.
    \item[] Guidelines:
    \begin{itemize}
        \item The answer \answerNA{} means that the abstract and introduction do not include the claims made in the paper.
        \item The abstract and/or introduction should clearly state the claims made, including the contributions made in the paper and important assumptions and limitations. A \answerNo{} or \answerNA{} answer to this question will not be perceived well by the reviewers. 
        \item The claims made should match theoretical and experimental results, and reflect how much the results can be expected to generalize to other settings. 
        \item It is fine to include aspirational goals as motivation as long as it is clear that these goals are not attained by the paper. 
    \end{itemize}

\item {\bf Limitations}
    \item[] Question: Does the paper discuss the limitations of the work performed by the authors?
    \item[] Answer: \answerYes{} 
    \item[] Justification: We discuss the limitations of our work in Section~\ref{sec:limitation}.
    \item[] Guidelines:
    \begin{itemize}
        \item The answer \answerNA{} means that the paper has no limitation while the answer \answerNo{} means that the paper has limitations, but those are not discussed in the paper. 
        \item The authors are encouraged to create a separate ``Limitations'' section in their paper.
        \item The paper should point out any strong assumptions and how robust the results are to violations of these assumptions (e.g., independence assumptions, noiseless settings, model well-specification, asymptotic approximations only holding locally). The authors should reflect on how these assumptions might be violated in practice and what the implications would be.
        \item The authors should reflect on the scope of the claims made, e.g., if the approach was only tested on a few datasets or with a few runs. In general, empirical results often depend on implicit assumptions, which should be articulated.
        \item The authors should reflect on the factors that influence the performance of the approach. For example, a facial recognition algorithm may perform poorly when image resolution is low or images are taken in low lighting. Or a speech-to-text system might not be used reliably to provide closed captions for online lectures because it fails to handle technical jargon.
        \item The authors should discuss the computational efficiency of the proposed algorithms and how they scale with dataset size.
        \item If applicable, the authors should discuss possible limitations of their approach to address problems of privacy and fairness.
        \item While the authors might fear that complete honesty about limitations might be used by reviewers as grounds for rejection, a worse outcome might be that reviewers discover limitations that aren't acknowledged in the paper. The authors should use their best judgment and recognize that individual actions in favor of transparency play an important role in developing norms that preserve the integrity of the community. Reviewers will be specifically instructed to not penalize honesty concerning limitations.
    \end{itemize}

\item {\bf Theory assumptions and proofs}
    \item[] Question: For each theoretical result, does the paper provide the full set of assumptions and a complete (and correct) proof?
    \item[] Answer: \answerYes{} 
    \item[] Justification: We take an experimental approach to validate our hypothesis on redundancy in speech foundation model fine-tuned for deepfake detection. Details on the motivation and the experiments are presented in Section~\ref{sec:method} including the training objectives of our framework.
    \item[] Guidelines:
    \begin{itemize}
        \item The answer \answerNA{} means that the paper does not include theoretical results. 
        \item All the theorems, formulas, and proofs in the paper should be numbered and cross-referenced.
        \item All assumptions should be clearly stated or referenced in the statement of any theorems.
        \item The proofs can either appear in the main paper or the supplemental material, but if they appear in the supplemental material, the authors are encouraged to provide a short proof sketch to provide intuition. 
        \item Inversely, any informal proof provided in the core of the paper should be complemented by formal proofs provided in appendix or supplemental material.
        \item Theorems and Lemmas that the proof relies upon should be properly referenced. 
    \end{itemize}

    \item {\bf Experimental result reproducibility}
    \item[] Question: Does the paper fully disclose all the information needed to reproduce the main experimental results of the paper to the extent that it affects the main claims and/or conclusions of the paper (regardless of whether the code and data are provided or not)?
    \item[] Answer: \answerYes{} 
    \item[] Justification: Our proposed framework is described in detail in Section~\ref{sec:method}. We also provide details regarding its implementation in Section~\ref{sec:exp_setup}.
    \item[] Guidelines:
    \begin{itemize}
        \item The answer \answerNA{} means that the paper does not include experiments.
        \item If the paper includes experiments, a \answerNo{} answer to this question will not be perceived well by the reviewers: Making the paper reproducible is important, regardless of whether the code and data are provided or not.
        \item If the contribution is a dataset and\slash or model, the authors should describe the steps taken to make their results reproducible or verifiable. 
        \item Depending on the contribution, reproducibility can be accomplished in various ways. For example, if the contribution is a novel architecture, describing the architecture fully might suffice, or if the contribution is a specific model and empirical evaluation, it may be necessary to either make it possible for others to replicate the model with the same dataset, or provide access to the model. In general. releasing code and data is often one good way to accomplish this, but reproducibility can also be provided via detailed instructions for how to replicate the results, access to a hosted model (e.g., in the case of a large language model), releasing of a model checkpoint, or other means that are appropriate to the research performed.
        \item While NeurIPS does not require releasing code, the conference does require all submissions to provide some reasonable avenue for reproducibility, which may depend on the nature of the contribution. For example
        \begin{enumerate}
            \item If the contribution is primarily a new algorithm, the paper should make it clear how to reproduce that algorithm.
            \item If the contribution is primarily a new model architecture, the paper should describe the architecture clearly and fully.
            \item If the contribution is a new model (e.g., a large language model), then there should either be a way to access this model for reproducing the results or a way to reproduce the model (e.g., with an open-source dataset or instructions for how to construct the dataset).
            \item We recognize that reproducibility may be tricky in some cases, in which case authors are welcome to describe the particular way they provide for reproducibility. In the case of closed-source models, it may be that access to the model is limited in some way (e.g., to registered users), but it should be possible for other researchers to have some path to reproducing or verifying the results.
        \end{enumerate}
    \end{itemize}

\item {\bf Open access to data and code}
    \item[] Question: Does the paper provide open access to the data and code, with sufficient instructions to faithfully reproduce the main experimental results, as described in supplemental material?
    \item[] Answer: \answerYes{} 
    \item[] Justification: All datasets employed in this study are publicly available. We will provide the code for the final paper.
    \item[] Guidelines:
    \begin{itemize}
        \item The answer \answerNA{} means that paper does not include experiments requiring code.
        \item Please see the NeurIPS code and data submission guidelines (\url{https://neurips.cc/public/guides/CodeSubmissionPolicy}) for more details.
        \item While we encourage the release of code and data, we understand that this might not be possible, so \answerNo{} is an acceptable answer. Papers cannot be rejected simply for not including code, unless this is central to the contribution (e.g., for a new open-source benchmark).
        \item The instructions should contain the exact command and environment needed to run to reproduce the results. See the NeurIPS code and data submission guidelines (\url{https://neurips.cc/public/guides/CodeSubmissionPolicy}) for more details.
        \item The authors should provide instructions on data access and preparation, including how to access the raw data, preprocessed data, intermediate data, and generated data, etc.
        \item The authors should provide scripts to reproduce all experimental results for the new proposed method and baselines. If only a subset of experiments are reproducible, they should state which ones are omitted from the script and why.
        \item At submission time, to preserve anonymity, the authors should release anonymized versions (if applicable).
        \item Providing as much information as possible in supplemental material (appended to the paper) is recommended, but including URLs to data and code is permitted.
    \end{itemize}

\item {\bf Experimental setting/details}
    \item[] Question: Does the paper specify all the training and test details (e.g., data splits, hyperparameters, how they were chosen, type of optimizer) necessary to understand the results?
    \item[] Answer: \answerYes{} 
    \item[] Justification: We provide experimental details in Section~\ref{sec:exp_setup} and in Appendix~\ref{app:hyperparameters}.
    \item[] Guidelines:
    \begin{itemize}
        \item The answer \answerNA{} means that the paper does not include experiments.
        \item The experimental setting should be presented in the core of the paper to a level of detail that is necessary to appreciate the results and make sense of them.
        \item The full details can be provided either with the code, in appendix, or as supplemental material.
    \end{itemize}

\item {\bf Experiment statistical significance}
    \item[] Question: Does the paper report error bars suitably and correctly defined or other appropriate information about the statistical significance of the experiments?
    \item[] Answer: \answerNo{} 
    \item[] Justification: We did not report error bars or confidence intervals in the paper.
    \item[] Guidelines:
    \begin{itemize}
        \item The answer \answerNA{} means that the paper does not include experiments.
        \item The authors should answer \answerYes{} if the results are accompanied by error bars, confidence intervals, or statistical significance tests, at least for the experiments that support the main claims of the paper.
        \item The factors of variability that the error bars are capturing should be clearly stated (for example, train/test split, initialization, random drawing of some parameter, or overall run with given experimental conditions).
        \item The method for calculating the error bars should be explained (closed form formula, call to a library function, bootstrap, etc.)
        \item The assumptions made should be given (e.g., Normally distributed errors).
        \item It should be clear whether the error bar is the standard deviation or the standard error of the mean.
        \item It is OK to report 1-sigma error bars, but one should state it. The authors should preferably report a 2-sigma error bar than state that they have a 96\% CI, if the hypothesis of Normality of errors is not verified.
        \item For asymmetric distributions, the authors should be careful not to show in tables or figures symmetric error bars that would yield results that are out of range (e.g., negative error rates).
        \item If error bars are reported in tables or plots, the authors should explain in the text how they were calculated and reference the corresponding figures or tables in the text.
    \end{itemize}

\item {\bf Experiments compute resources}
    \item[] Question: For each experiment, does the paper provide sufficient information on the computer resources (type of compute workers, memory, time of execution) needed to reproduce the experiments?
    \item[] Answer: \answerYes{} 
    \item[] Justification: We provided our configuration in Appendix~\ref{app:hyperparameters}
    \item[] Guidelines:
    \begin{itemize}
        \item The answer \answerNA{} means that the paper does not include experiments.
        \item The paper should indicate the type of compute workers CPU or GPU, internal cluster, or cloud provider, including relevant memory and storage.
        \item The paper should provide the amount of compute required for each of the individual experimental runs as well as estimate the total compute. 
        \item The paper should disclose whether the full research project required more compute than the experiments reported in the paper (e.g., preliminary or failed experiments that didn't make it into the paper). 
    \end{itemize}
    
\item {\bf Code of ethics}
    \item[] Question: Does the research conducted in the paper conform, in every respect, with the NeurIPS Code of Ethics \url{https://neurips.cc/public/EthicsGuidelines}?
    \item[] Answer: \answerYes{} 
    \item[] Justification: We have reviewed and respect NeurIPS Code of Ethics.
    \item[] Guidelines:
    \begin{itemize}
        \item The answer \answerNA{} means that the authors have not reviewed the NeurIPS Code of Ethics.
        \item If the authors answer \answerNo, they should explain the special circumstances that require a deviation from the Code of Ethics.
        \item The authors should make sure to preserve anonymity (e.g., if there is a special consideration due to laws or regulations in their jurisdiction).
    \end{itemize}

\item {\bf Broader impacts}
    \item[] Question: Does the paper discuss both potential positive societal impacts and negative societal impacts of the work performed?
    \item[] Answer: \answerYes{} 
    \item[] Justification:  As described in Section~\ref{sec:introduction}, our proposed framework is designed to detect deepfake audios, hence our work can be used for positive societal impacts.
    \item[] Guidelines:
    \begin{itemize}
        \item The answer \answerNA{} means that there is no societal impact of the work performed.
        \item If the authors answer \answerNA{} or \answerNo, they should explain why their work has no societal impact or why the paper does not address societal impact.
        \item Examples of negative societal impacts include potential malicious or unintended uses (e.g., disinformation, generating fake profiles, surveillance), fairness considerations (e.g., deployment of technologies that could make decisions that unfairly impact specific groups), privacy considerations, and security considerations.
        \item The conference expects that many papers will be foundational research and not tied to particular applications, let alone deployments. However, if there is a direct path to any negative applications, the authors should point it out. For example, it is legitimate to point out that an improvement in the quality of generative models could be used to generate Deepfakes for disinformation. On the other hand, it is not needed to point out that a generic algorithm for optimizing neural networks could enable people to train models that generate Deepfakes faster.
        \item The authors should consider possible harms that could arise when the technology is being used as intended and functioning correctly, harms that could arise when the technology is being used as intended but gives incorrect results, and harms following from (intentional or unintentional) misuse of the technology.
        \item If there are negative societal impacts, the authors could also discuss possible mitigation strategies (e.g., gated release of models, providing defenses in addition to attacks, mechanisms for monitoring misuse, mechanisms to monitor how a system learns from feedback over time, improving the efficiency and accessibility of ML).
    \end{itemize}
    
\item {\bf Safeguards}
    \item[] Question: Does the paper describe safeguards that have been put in place for responsible release of data or models that have a high risk for misuse (e.g., pre-trained language models, image generators, or scraped datasets)?
    \item[] Answer: \answerNA{} 
    \item[] Justification: The paper poses no such risks.
    \item[] Guidelines:
    \begin{itemize}
        \item The answer \answerNA{} means that the paper poses no such risks.
        \item Released models that have a high risk for misuse or dual-use should be released with necessary safeguards to allow for controlled use of the model, for example by requiring that users adhere to usage guidelines or restrictions to access the model or implementing safety filters. 
        \item Datasets that have been scraped from the Internet could pose safety risks. The authors should describe how they avoided releasing unsafe images.
        \item We recognize that providing effective safeguards is challenging, and many papers do not require this, but we encourage authors to take this into account and make a best faith effort.
    \end{itemize}

\item {\bf Licenses for existing assets}
    \item[] Question: Are the creators or original owners of assets (e.g., code, data, models), used in the paper, properly credited and are the license and terms of use explicitly mentioned and properly respected?
    \item[] Answer: \answerYes{} 
    \item[] Justification: The employed datasets, code, and model backbones are all properly cited (Section~\ref{sec:exp_setup}).
    \item[] Guidelines:
    \begin{itemize}
        \item The answer \answerNA{} means that the paper does not use existing assets.
        \item The authors should cite the original paper that produced the code package or dataset.
        \item The authors should state which version of the asset is used and, if possible, include a URL.
        \item The name of the license (e.g., CC-BY 4.0) should be included for each asset.
        \item For scraped data from a particular source (e.g., website), the copyright and terms of service of that source should be provided.
        \item If assets are released, the license, copyright information, and terms of use in the package should be provided. For popular datasets, \url{paperswithcode.com/datasets} has curated licenses for some datasets. Their licensing guide can help determine the license of a dataset.
        \item For existing datasets that are re-packaged, both the original license and the license of the derived asset (if it has changed) should be provided.
        \item If this information is not available online, the authors are encouraged to reach out to the asset's creators.
    \end{itemize}

\item {\bf New assets}
    \item[] Question: Are new assets introduced in the paper well documented and is the documentation provided alongside the assets?
    \item[] Answer: \answerNA{} 
    \item[] Justification: The paper does not introduce new assets.
    \item[] Guidelines:
    \begin{itemize}
        \item The answer \answerNA{} means that the paper does not release new assets.
        \item Researchers should communicate the details of the dataset\slash code\slash model as part of their submissions via structured templates. This includes details about training, license, limitations, etc. 
        \item The paper should discuss whether and how consent was obtained from people whose asset is used.
        \item At submission time, remember to anonymize your assets (if applicable). You can either create an anonymized URL or include an anonymized zip file.
    \end{itemize}

\item {\bf Crowdsourcing and research with human subjects}
    \item[] Question: For crowdsourcing experiments and research with human subjects, does the paper include the full text of instructions given to participants and screenshots, if applicable, as well as details about compensation (if any)? 
    \item[] Answer: \answerNA{} 
    \item[] Justification: The paper does not involve crowdsourcing nor research with human subjects.
    \item[] Guidelines:
    \begin{itemize}
        \item The answer \answerNA{} means that the paper does not involve crowdsourcing nor research with human subjects.
        \item Including this information in the supplemental material is fine, but if the main contribution of the paper involves human subjects, then as much detail as possible should be included in the main paper. 
        \item According to the NeurIPS Code of Ethics, workers involved in data collection, curation, or other labor should be paid at least the minimum wage in the country of the data collector. 
    \end{itemize}

\item {\bf Institutional review board (IRB) approvals or equivalent for research with human subjects}
    \item[] Question: Does the paper describe potential risks incurred by study participants, whether such risks were disclosed to the subjects, and whether Institutional Review Board (IRB) approvals (or an equivalent approval/review based on the requirements of your country or institution) were obtained?
    \item[] Answer: \answerNA{} 
    \item[] Justification:  The paper does not involve crowdsourcing nor research with human subjects.
    \item[] Guidelines:
    \begin{itemize}
        \item The answer \answerNA{} means that the paper does not involve crowdsourcing nor research with human subjects.
        \item Depending on the country in which research is conducted, IRB approval (or equivalent) may be required for any human subjects research. If you obtained IRB approval, you should clearly state this in the paper. 
        \item We recognize that the procedures for this may vary significantly between institutions and locations, and we expect authors to adhere to the NeurIPS Code of Ethics and the guidelines for their institution. 
        \item For initial submissions, do not include any information that would break anonymity (if applicable), such as the institution conducting the review.
    \end{itemize}

\item {\bf Declaration of LLM usage}
    \item[] Question: Does the paper describe the usage of LLMs if it is an important, original, or non-standard component of the core methods in this research? Note that if the LLM is used only for writing, editing, or formatting purposes and does \emph{not} impact the core methodology, scientific rigor, or originality of the research, declaration is not required.
    \item[] Answer: \answerNA{} 
    \item[] Justification: LLMs were only used for language editing and formatting and did not impact the core methodology.
    \item[] Guidelines:
    \begin{itemize}
        \item The answer \answerNA{} means that the core method development in this research does not involve LLMs as any important, original, or non-standard components.
        \item Please refer to our LLM policy in the NeurIPS handbook for what should or should not be described.
    \end{itemize}

\end{enumerate}

\end{document}